\def\USETWOCOLUMNS{1} %
\def\USEARXIV{1} %

\if\USETWOCOLUMNS1
\documentclass[twoside,twocolumn,9pt]{article} 
\else
\documentclass[12pt]{article} 
\fi
\usepackage{extsizes}
\usepackage[super,sort&compress,comma]{natbib} 
\usepackage[version=3]{mhchem}
\usepackage[left=1.5cm, right=1.5cm, top=1.785cm, bottom=2.0cm]{geometry}
\usepackage{balance}
\usepackage{mathptmx}
\usepackage{sectsty}
\usepackage{graphicx} 
\usepackage{lastpage}
\usepackage[format=plain,justification=justified,singlelinecheck=false,font={stretch=1.125,small,sf},labelfont=bf,labelsep=space]{caption}
\usepackage{float}
\usepackage{fancyhdr}
\usepackage{fnpos}
\usepackage[english]{babel}
\addto{\captionsenglish}{%
  \renewcommand{\refname}{Notes and references}
}
\usepackage{array}
\usepackage{droidsans}
\usepackage{charter}
\usepackage[T1]{fontenc}
\usepackage[usenames,dvipsnames]{xcolor}
\usepackage{setspace}
\usepackage[compact]{titlesec}
\usepackage{hyperref}

\definecolor{cream}{RGB}{222,217,201}

\usepackage{amssymb}
\usepackage{pdflscape} %
\graphicspath{{pics}}
\usepackage{dcolumn}
\usepackage{booktabs}
\usepackage{longtable}
\usepackage[textsize=small,textwidth=2.5cm]{todonotes}
\usepackage{units}
\usepackage{xspace}
\let\oldtheequation\theequation
\makeatletter
\def\tagform@#1{\maketag@@@{\ignorespaces#1\unskip\@@italiccorr}}
\renewcommand{\theequation}{(\oldtheequation)}
\makeatother 

\addto\extrasenglish{%
}

\newcommand{\icm}{cm^{-1}}

\newcommand{\ket}[1]{\ensuremath{  | {#1} \rangle}}

\newcommand{\lit}[1]{Ref.~\mbox{[\!\!\citenum{#1}]}\xspace}
\newcommand{\lits}[1]{Refs.~\mbox{[\!\!\citenum{#1}]}\xspace}

\begin{document}
\pagestyle{fancy}
\thispagestyle{plain}
\fancypagestyle{plain}{
\renewcommand{\headrulewidth}{0pt}
}
\makeFNbottom
\makeatletter
\renewcommand\LARGE{\@setfontsize\LARGE{15pt}{17}}
\renewcommand\Large{\@setfontsize\Large{12pt}{14}}
\renewcommand\large{\@setfontsize\large{10pt}{12}}
\renewcommand\footnotesize{\@setfontsize\footnotesize{7pt}{10}}
\makeatother

\renewcommand{\thefootnote}{\fnsymbol{footnote}}
\renewcommand\footnoterule{\vspace*{1pt}%
\color{cream}\hrule width 3.5in height 0.4pt \color{black}\vspace*{5pt}} 
\setcounter{secnumdepth}{5}

\makeatletter 
\renewcommand\@biblabel[1]{#1}            
\renewcommand\@makefntext[1]%
{\noindent\makebox[0pt][r]{\@thefnmark\,}#1}
\makeatother 
\renewcommand{\figurename}{\small{Fig.}~}
\sectionfont{\sffamily\Large}
\subsectionfont{\normalsize}
\subsubsectionfont{\bf}
\setstretch{1.125} %
\setlength{\skip\footins}{0.8cm}
\setlength{\footnotesep}{0.25cm}
\setlength{\jot}{10pt}
\titlespacing*{\section}{0pt}{4pt}{4pt}
\titlespacing*{\subsection}{0pt}{15pt}{1pt}
\fancyfoot{}
\fancyfoot[LO,RE]{\vspace{-7.1pt}\includegraphics[height=9pt]{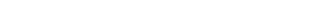}}
\fancyfoot[CO]{\vspace{-7.1pt}\hspace{11.9cm}\includegraphics{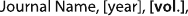}}
\fancyfoot[CE]{\vspace{-7.2pt}\hspace{-13.2cm}\includegraphics{head_foot/RF}}
\fancyfoot[RO]{\footnotesize{\sffamily{1--\pageref{LastPage} ~\textbar  \hspace{2pt}\thepage}}}
\fancyfoot[LE]{\footnotesize{\sffamily{\thepage~\textbar\hspace{4.65cm} 1--\pageref{LastPage}}}}
\fancyhead{}
\renewcommand{\headrulewidth}{0pt} 
\renewcommand{\footrulewidth}{0pt}
\setlength{\arrayrulewidth}{1pt}
\setlength{\columnsep}{6.5mm}
\setlength\bibsep{1pt}
\makeatletter 
\newlength{\figrulesep} 
\setlength{\figrulesep}{0.5\textfloatsep} 

\newcommand{\topfigrule}{\vspace*{-1pt}%
\noindent{\color{cream}\rule[-\figrulesep]{\columnwidth}{1.5pt}} }

\newcommand{\botfigrule}{\vspace*{-2pt}%
\noindent{\color{cream}\rule[\figrulesep]{\columnwidth}{1.5pt}} }

\newcommand{\dblfigrule}{\vspace*{-1pt}%
\noindent{\color{cream}\rule[-\figrulesep]{\textwidth}{1.5pt}} }

\makeatother
\if\USETWOCOLUMNS1
\twocolumn[
  \begin{@twocolumnfalse}
\fi
{\includegraphics[height=30pt]{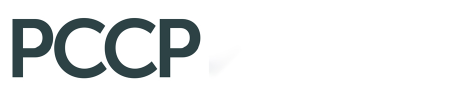}\hfill\raisebox{0pt}[0pt][0pt]{\includegraphics[height=55pt]{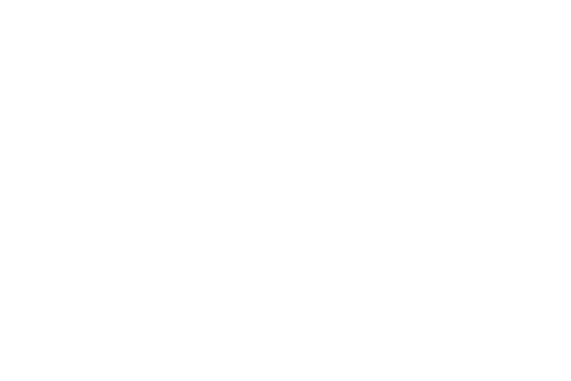}}\\[1ex]
\includegraphics[width=18.5cm]{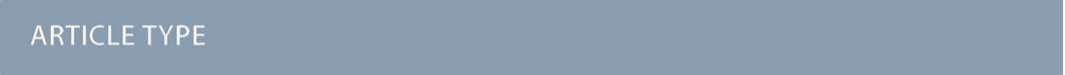}}\par
\vspace{1em}
\sffamily
\begin{tabular}{m{4.5cm} p{13.5cm} }

\includegraphics{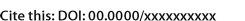} & \noindent\LARGE{\textbf{2500 vibronic eigenstates of the \ce{NO3} radical}} \\%
\vspace{0.3cm} & \vspace{0.3cm} \\

 & \noindent\large{Henrik R.~Larsson,$^{\ast}$\textit{$^{a}$} Alexandra Viel\textit{$^{b}$}} \\

\includegraphics{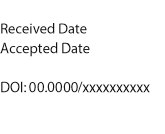} & \noindent\normalsize{
The nitrate radical \ce{NO3} plays an important role in atmospheric chemistry, yet many aspects of its coupled and anharmonic vibronic structure remain elusive.
Here, using an accurate, coupled full-dimensional diabatic potential that includes five 
electronic states, we revisit the vibronic spectrum associated with the electronic $\tilde X ^2A_2'$ state.
Using recently developed tensor network state methods, we are able to compute more than 2500 vibronic states, thereby increasing the number of computed full-dimensional states by a factor of 50, compared to previous work.
While we obtain good agreement with experiment for most of the assigned vibronic levels, for several others, we observe striking disagreement.
Further, for the antisymmetric bending motion
we find remarkably large symmetry-induced level splittings that are larger than the zero-order reference.
We discuss non-negligible
nonadiabatic effects and show that the Born-Oppenheimer approximation %
leads to significant errors in the spectrum.
} \\%

\end{tabular}

\if\USETWOCOLUMNS1
 \end{@twocolumnfalse} \vspace{0.6cm}
  ]
\fi
\renewcommand*\rmdefault{bch}\normalfont\upshape
\rmfamily
\section*{}
\vspace{-1cm}

\footnotetext{\textit{$^{a}$~Department of Chemistry and Biochemistry, University of California, Merced, CA 95343}; E-mail: NO3a[at]larsson-research.$\delta$e}
\footnotetext{\textit{$^{b}$~Univ.~Rennes, CNRS, IPR (Institut de Physique de Rennes) - UMR 6251,
F-35000 Rennes, France. E-mail: alexandra.viel[at]univ-rennes1.fr}}

\footnotetext{\dag~Electronic Supplementary Information (ESI) available: Additional infrared spectra. See DOI: 10.1039/cXCP00000x/}
\section{Introduction}
\label{sec:intro}

\ce{NO3} is one of the first scientifically characterized radicals and has a century-old scientific history rich of drama.\cite{%
Etude1882chappuis,
Uber1907warburg,
Absorption1937jones,
Infrared1965guillory,
Erratum1967morris,
Distortions1970olsena,
Infrared1985ishiwata,
Examination1991weaver,
Potential1991stanton,
Choice1992stanton,
Detailed2000eisfeld,
Theoretical2003eisfeld,
N32012simmons,
No2012grubb,
Vibrational2015hirota,
Assignment2018hirota,
Simulation2022stanton,
Simulation2022williams}
\ce{NO3} is important 
in atmospheric nighttime chemistry,\cite{Nitrate1991wayne,Gasphase2005monks,Nighttime2012brown}
and, under polluted conditions in smoggy urban environments, also in atmospheric daytime chemistry.\cite{Direct2003geyer,Gasphase2005monks,Nighttime2012brown}
Like many other small molecules and radicals,\cite{Symmetry1991kaldor,Beryllium2009merritt,Elusive2009patkowski,Spectroscopic2014sharma,Minimal2020larsson,Variational2020wang,Matrix2022larsson,Chromium2022larsson,Spectroscopy2023neumark}
\ce{NO3} has both an intriguing electronic and vibronic structure.
Throughout the decades, many scientists hence investigated various aspects of \ce{NO3}; see %
\lits{%
Etude1882chappuis,
Uber1907warburg,
Absorption1937jones,
Infrared1965guillory,
Erratum1967morris,
Distortions1970olsena,
Infrared1985ishiwata,
Examination1991weaver,
Potential1991stanton,
Choice1992stanton,
Detailed2000eisfeld,
Theoretical2003eisfeld,
N32012simmons,
No2012grubb,
Vibrational2015hirota,
Assignment2018hirota,
Simulation2022stanton,
Simulation2022williams,
Nitrate1991wayne,
Direct2003geyer,Gasphase2005monks,Nighttime2012brown,
Infrared1965cramarossa,
Photochemistry1978graham,
Absorption1982marinelli,
Absorption1985burrows,
Temperature1986sander,
Fourier1987friedl,
Temperature1987cantrell,
Vibronic1991hirota,
NO31992kim,
Ground1994mayer,
Temperature1994yokelson,
Nearinfrared1997hirota,
Infrared1998kawaguchi,
Initio2001eisfeld,
Visible2003orphal,
Cavity2005deev,
New2005okumura,
Vibronic2007stanton,
Higherorder2008faraji,
Photoionizationinduced2008viel,
Conflicting2009beckers,
Vibronic2009stantona,
Vibronic2009stanton,
Fourier2011kawaguchi,
Global2012xiao,
Quasiclassical2013fu,
Fourier2013fujimori,
FTIR2013kawaguchi,
Spectroscopic2013takematsu,
Fulldimensional2014eisfeld,
Communication2014homayoon,
Jet2015codd,
Vibronic2017eisfeld,
Infrared2017kawaguchi,
Initio2017mukherjee,
Infrared2018kawaguchia,
Infrared2018kawaguchi,
NO32018viel,
Neural2018williams,
Quantum2019weike,
Nature2020kalemos,
HighResolution2020babin,
Infrared2021kawaguchi,
Accurate2021viel,
Laser2022fukushima,
Infrared2022kawaguchi,
Spectroscopy2023neumark,
Molecules2024sharma}
for a selected list of examples.
Despite the impressive amount of both experimental and computational research on \ce{NO3},
many aspects of \ce{NO3} remain elusive. Here specifically, we focus on the mysteries of the vibronic eigenspectrum of levels associated with the $\tilde{X} ^2A_2'$ electronic ground state.

The vibrations of \ce{NO3} are strongly anharmonic and many of the vibrational motions are coupled. 
Further, 
the  $\tilde{X} ^2A_2'$ state is coupled to the degenerate $\tilde{A} ^2E''$, and $\tilde{B} ^2E'$  excited states through pseudo Jahn-Teller and dynamical pseudo Jahn-Teller effects. 
These excited states themselves are subject to a strong Jahn-Teller $e \times E$ effect.
Thus, a \emph{complete} description of all aspects of \ce{NO3} requires 
the inclusion of these four $\tilde A$ and $\tilde B$ states  
in the computation of accurate vibronic eigenstates even for the $\tilde{X}$ ground state.

Several theorists computed the vibrational and/or vibronic 
eigenspectrum using different potential energy surfaces (PESs).
Using a qualitative vibronic coupling model,\cite{Multimode1984koppel} and neglecting the umbrella motion, Stanton computed vibronic eigenstates and hinted at a possible reassignment of one of the fundamentals.\cite{Vibronic2007stanton}
Later, he improved this model to  perform more accurate simulations in reduced dimensionality that led to mounting evidence for a required reassignment of the fundamental.\cite{N32012simmons,Simulation2022stanton}
The improved model also 
helped to analyze the photodetachment spectrum of \ce{NO3-}.\cite{HighResolution2020babin}
In addition, using a quartic force field, Stanton computed the vibrational spectrum within the Born-Oppenheimer approximation, including the infrared (IR) intensities.\cite{Vibronic2009stantona}
Homayoon and Bowman, using a PES from Morokuma et al.,\cite{Global2012xiao}
performed Born-Oppenheimer full-dimensional vibrational eigenstate computations using a semi-empirical coordinate scaling and could confirm the reassignment.\cite{Communication2014homayoon}
Viel and Eisfeld,
using a full-dimensional vibronic coupling model with higher-order couplings terms, obtained vibronic levels and IR spectra
that, despite a more flexible PES model compared to the previously mentioned PESs, led to no improved agreement with experiment, compared to the previous computations. 
However, they could confirm some previous results and revealed several non-adiabatic effects.\cite{NO32018viel}
Later, Viel and Eisfeld, joined by Williams, presented a completely new PES still based on the vibronic coupling model but now with coordinate-depending parameters fitted using artificial neural networks.\cite{Neural2018williams}
Initial Born-Oppenheimer computations for selected vibrational states  showed much improved agreement with experiment, compared to the PES previously developed by the two groups.\cite{Diabatic2019williams}
Subsequently, Viel et al.~used the new PES to simulate the photodetachment spectra of \ce{NO3-} that, among others, led to the understanding of  a hitherto unexplored hot band in the experimental spectrum.\cite{Accurate2021viel,Simulation2022williams}

Despite the aforementioned body of work on the vibrational/vibronic spectrum, many simulations were limited by, e.g.,
the use of semi-empirical parameters, PESs that are not flexible enough,
computations done in reduced dimensionality, or the neglect of non-Born-Oppenheimer effects.
In addition, 
so far the largest number of full-dimensional vibronic states computed was 52,\cite{NO32018viel}
and most of the research focused on the eigenstates below $\unit[2000]{\icm}$. Undeniably, this low-energy region is the most important one in the IR spectrum. However, more experimental vibronic states have been assigned with energies around $\unit[3000]{\icm}$ that so far lack an accurate simulated counterpart.\footnote{Stanton's initial study assigned states up to $\unit[2500]{\icm}$ and some higher-lying states,\cite{Vibronic2007stanton} but these levels are only qualitatively correct and do not include the umbrella motion.}
In addition, high-energy vibronic states reveal more details about non-adiabatic coupling as these, in terms of energy, are closer to the next electronic state.
Many mysteries about the vibronic spectrum hence remain and hint at possible misassigned experimental states.\cite{Simulation2022stanton}
Driven by these motivations, here we report the accurate computation of more than 2500 eigenstates. This increases the number of 
rigorously computed vibronic states for this system by a factor of $~50$.
Moreover, we assign all states up to $\unit[3000]{\icm}$, thereby increasing the number of assigned full-dimensional states by a  factor of $\sim 5$.

This work is possible by two recent methodological advancements.
The first advancement is the development of the aforementioned PES based on a vibronic coupling model and an intricate artificial neural network.\cite{Neural2018williams,Diabatic2019williams,Accurate2021viel}
The second advancement is the very efficient and accurate computation of vibrational states using the density matrix renormalization group (DMRG)
and tree tensor network states (TTNS).\cite{Computing2019larsson,Tensor2024larsson}
This approach recently enabled the computation of the vibrational eigenspectrum of the 15-dimensional fluxional Zundel ion.\cite{Stateresolved2022larsson}
Other groups demonstrated a similar efficiency of related tensor network methods targeting vibrational spectra; see
\lits{Automated2015meier,Calculating2016rakhuba,Flexible2023glaser,Eigenstate2024hoppe,Using2024wodraszka} 
for some selected recent examples.
Here, we extend our TTNS approach to the computation of vibronic spectra and present an accurate way for refitting the diabatic potential matrix into a tensor network form. We further test this approach using a complementary method based on pruning.\cite{Efficient2016larsson,Dynamical2017larsson,Resonance2018larsson}

Our outline is as follows: \autoref{sec:methods} introduces more details about 
the used coordinate system and
the employed simulation methods. 
\autoref{sec:setup} specifies details about the simulation parameters and \autoref{sec:setup_assignment} introduces our methods for assigning the vibronic eigenstates. The results are presented and discussed in \autoref{sec:results}.
\autoref{subsec:results_overview} discusses the overall eigenspectrum, 
\autoref{subsec:assignment} focuses on the assigned eigenstates, 
\autoref{subsec:IR_spec} deals with the infrared spectrum, and 
\autoref{sec:nonadiabatic_effects} discusses non-adiabatic effects.
We conclude in \autoref{sec:conclusions}.

\section{Simulation methods}
\label{sec:methods}

We use the curvilinear coordinates that are 
adapted to the $C_{3v}$ symmetry of \ce{NO3} and
based on Radau and hyperspherical coordinates
as defined in \lits{Quantum2007evenhuis,Photoionizationinduced2008viel},
and we use a  total angular momentum quantum number of $J=0$.
See also  \lits{Orthogonal2009ragni,Umbrella2016ragni}
for a similar type of coordinate system.
The corresponding  kinetic energy operator contains terms that are not of the form of sum of products of one-dimensional operators.
To use a sum-of-product operator, 
in \lits{Quantum2007evenhuis,Photoionizationinduced2008viel}, 
second- and fourth-order Taylor expansions of the problematic terms near $C_{3v}$-symmetric configurations were provided.
The fourth-order expansion was tested on the \ce{NO3} radical in \lit{Vibronic2017eisfeld}.
There, vibronic energy levels supported by the $E''$ state computed with an alternative coordinate system and its exact associated kinetic operator were found in very good agreement with the ones computed with the curvilinear coordinates and the approximated kinetic energy operator.
This demonstrates that the fourth-order approximation is excellent, even for geometries away from the $C_{3v}$ reference that are imposed by the triple-well structure of the lower $E''$ adiabatic surface.
Accordingly, we use the fourth-order approximation in the following.
The quasi-exact kinetic energy  operator then can be described as sum over  58 product terms.

The diabatic PESs are described by the functional form introduced  in \lits{Neural2018williams,Diabatic2019williams,Accurate2021viel}.
The coupled PESs represent the five lowest electronic energies  
of \ce{NO3}, namely the $\tilde{X} ^2A_2'$, $\tilde{A} ^2E''$, and $\tilde{B} ^2E'$ electronic states.\cite{Accurate2021viel}
The $E$ states are doubly degenerate.
The PES is based on accurate reference data at the multi-reference configuration interaction level using an adapted basis set of triple-$\zeta$ quality; see \lit{Neural2018williams} and references therein for further information.
The diabatic potential energy  matrix is of the form of 
  \begin{equation}
  \hat V = 
    \begin{pmatrix}
      \hat V_{11} & \hat V_{12} & \hat V_{13} & \hat V_{14} & \hat V_{15}\\
      \hat V_{21} & \hat V_{22} & \hat V_{23} & \hat V_{24} & \hat V_{25}\\
      \hat V_{31} & \hat V_{32} & \hat V_{33} & \hat V_{34} & \hat V_{35}\\
      \hat V_{41} & \hat V_{42} & \hat V_{43} & \hat V_{44} & \hat V_{45}\\
      \hat V_{51} & \hat V_{52} & \hat V_{53} & \hat V_{54} & \hat V_{55}\\
    \end{pmatrix},
    \label{eq:vmatrix}
  \end{equation}
  where each operator $\hat V_{ij}$ is represented as function over all six coordinates. 
  The full-dimensional $5\times 5$ matrix  in \autoref{eq:vmatrix}
  results from an elaborated combination of diabatization by ansatz and artificial neural network fitting.\cite{Neural2018williams,Diabatic2019williams}
The coordinates used for the diabatic PES are not the aforementioned curvilinear coordinates but symmetry-adapted coordinates as described in the appendix of \lit{Accurate2021viel}. 
The part of the model containing
the artificial neural network depends on nine invariants that are explicitly given in that appendix, also. 
The symmetry-adapted coordinates are mandatory to encode the symmetry of the diabatic elements, while the nine invariants (\textit{i.e.}, totally symmetric coordinate polynomials) correspond to all unique possible combinations up to third order. They enable the flexibility of the neural network outputs and hence are responsible for the accuracy of the fit.

Recently, 
this PES has been used to model the photodetachment spectra of \ce{NO3-} both in the lower energy region\cite{Accurate2021viel} and in the energy range of the second excited state.\cite{Simulation2022williams}
The agreement with the experimental photodetachment spectra underlined the accuracy and the physical soundness of this PES.
The minimum of the ground-state adiabatic PES corresponds to a $D_{3h}$ symmetry, whereas the first and second excited states are strongly influenced by the Jahn-Teller $e \times E$ effect.
The PES model also contains the pseudo Jahn-Teller and dynamical pseudo Jahn-Teller couplings between the $\tilde{X} ^2A_2'$ and the two  $\tilde{A} ^2E''$, and $\tilde{B} ^2E'$  excited states.
The adiabatic representation of the PES can be described by an anharmonic ``bowl''  for the  electronic ground state, followed by two pairs of PESs containing conical intersections surrounded by three separate wells.

To compute the eigenstates, we use the TTNS approach developed in \lit{Computing2019larsson}.
TTNSs are the wavefunction ansatz of both the multilayer multiconfiguration time-dependent Hartree (ML-MCTDH) method\cite{Multilayer2003wang,Multilayer2008manthe,Multilayer2015wang,Wavepacket2017manthe}
and of extended versions of the density matrix renormalization group (DMRG).\cite{Densitymatrix2011schollwock,Practical2014orus,Density2020baiardi}
They allow for a very economic description of high-dimensional eigenstates.\cite{Low2012chan,Tensor2024larsson}
Here, the eigenstates will be computed using the DMRG sweep algorithm.\cite{Density1992white,Densitymatrix1993white}
We have previously demonstrated the accuracy and efficiency of  our TTNS-based DMRG approach by computing about 1000 eigenstates of the 15-dimensional fluxional Zundel ion to accuracies that are below the error of the PES fit.\cite{Stateresolved2022larsson}
The ML-MCTDH and DMRG  methods work most optimal if the Hamiltonian has a form that is compatible with TTNSs.
Many (but not all\cite{Nonhierarchical2024ellerbrock})
ML-MCTDH method users typically re-fit the potential as sum-of-product form.\cite{Product1996jackle,Transforming2020schroder}
Here, we found such forms to be less accurate (see \autoref{subsec:results_overview}) and   instead decompose the analytical form of the PES using matrix product operators (MPOs), which are straightforward extensions of matrix product states (MPSs), a subset of TTNSs, to operators and which straightforwardly  can be obtained numerically using singular value decomposition.\cite{Practical2014orus,Calculating2016rakhuba,Block22023zhai,Tensor2024larsson,Encoding2024hino}
Such a fit
can be viewed as alternative to the ``multilayer potfit'' approach.\cite{Multilayer2014otto}

To provide an alternative means of testing the accuracy of the TTNS method and the MPO decomposition, we use the previously developed dynamically pruned  (DP) discrete variable representation (DVR) approach (DP-DVR).\cite{Efficient2016larsson,PhaseSpace2018tannor} 
This is based on using only those configurations of the state in coordinate space that are larger than a certain wave amplitude threshold.
Phase-space localized bases are more sparse than coordinate-space localized bases,\cite{Quantum2004poiriera,Neumann2014shimshovitz,Using2015brown,pvb_math_tannor_2016,PhaseSpace2018tannor} but we showed that due to the diagonality of the potential operator in DVR representation, using DVRs is more efficient despite the need for storing more important configurations.\cite{Efficient2016larsson,PhaseSpace2018tannor}
Previously, one of us demonstrated the efficiency of the DP-DVR method for
vibrational resonance decay dynamics,\cite{Resonance2018larsson} strong-field electron dynamics,\cite{Comparing2018esry,Control2021larsson}
and non-adiabatic and photodissociation dynamics in combination with the MCTDH method (DP-MCTDH).\cite{Dynamical2017larsson}

\section{Setup of the computations}
\label{sec:setup}

The states were represented using sinc DVRs\cite{Novel1992colbert,Discretevariable2000light,PhaseSpace2018tannor} 
in each of the curvilinear coordinates.\footnote{We use an atomic mass of $\unit[15.9949138012]{Da}$ for \ce{^{16}O}  and $\unit[14.0030732884]{Da}$ for  \ce{^{14}N}.}
The basis-set parameters are given in \autoref{tab:basisset}.
These parameters were determined by comparing separate eigenstate computations for different coordinate ranges and basis sizes and by observing the reduced densities in position and momentum space. 
The parameters are similar to those from previous studies.\cite{NO32018viel,Simulation2022williams}
The TTNS and the MPO decompositions depend on one convergence parameter that specifies the size of the tensors, the so-called bond dimension $D$ (also known as rank, renormalized basis size, or number of single-particle functions in ML-MCTDH\cite{Tensor2024larsson}).
The diabatic potential matrix elements in \autoref{eq:vmatrix} were represented as MPO with two different maximal bond dimensions for two individual simulations analyzed in \autoref{sec:results}.
For the analysis of the first 180 states with excitation energies up to $\tilde E = \unit[3000]{\icm}$ we used an MPO with $D=1,800$,
whereas for the analysis of all $\sim 2500$ eigenstates (including the first 180), we used an MPO with $D=600$.
The $D=600$ MPO led to an accurate description of most but not all subtleties of the vibronic states that are reported below.

\begin{table}
\caption{Used basis-set parameters. The coordinate range is displayed in either bohr or radians, depending on the type of coordinate.}
\label{tab:basisset}
 \begin{tabular*}{.48\textwidth}{@{\extracolsep{\fill}}crrr}
    \toprule
    symbol & coordinate range & basis size\\
    \midrule
    $\rho$ & $[638, 770]$ & $56$ \\
    $\vartheta$ & $[0.845, 1.065]$ & $42$ \\
    $\varphi$ &  $[0.665, 0.905]$ & $48$\\
    $\theta$ & $[1.451, 1.691]$& $46$ \\
    $\phi$ &  $[0.817, 1.227]$& $40$\\
    $\chi$ &  $[2.734, 3.548]$& $26$\\
    \bottomrule
  \end{tabular*}
  \end{table}

We represented the eigenstates as MPS, a subset of TTNSs, using the following order of coordinates: electronic, $\rho$, $\vartheta$, $\phi$, $\theta$, $\varphi$, $\chi$. 
In some test computations,
we found this ordering to be optimal with respect to the bond dimension.
Due to the low dimensionality of the system, a tree structure is not required and hence we simply use a linear structure, as given by an MPS.
The MPS was used together with a maximal bond dimension of $30$. 
Renormalized basis states with singular values below $10^{-8}$ were not included, which means that the actual bond dimension for each tensor can be smaller than $30$.\cite{Controlling2003legeza,Computing2019larsson}
For the DMRG sweeps a relative convergence tolerance of $10^{-6}$ was used.
We tested the convergence of these parameters  through separate computations with tighter parameters. %
By shifting\cite{Iterative1973shavitt,CheMPS22014wouters} previously computed states up in energy, the eigenstates were computed one by one as described in  \lit{Computing2019larsson}. 

The accuracy of the MPO fit of the diabatic potentials was checked by computing the direct-product-DVR amplitudes of some TTNSs and pruning them using the DP-DVR procedure described in \lit{Efficient2016larsson} and a wave amplitude threshold of $5\cdot 10^{-7}$.
The smaller the threshold, the more accurate the representation of the DP-DVR eigenstates.
The used threshold led to energy errors that are on the order of $\unit[\sim2]{\icm}$, which is small enough for this accuracy check. 
The energies of the DP-DVR states were computed using the original and thus non-MPO-approximated potential surfaces and an algorithm for efficiently computing matrix-vector products described in \lit{Efficient2016larsson}. 
The IR stick spectra were generated using the adiabatic dipole moment surface (DMS) from \lit{NO32018viel}, which was converted to the diabatic representation to compute dipole matrix elements. 
The adiabatic-to-diabatic transformation was obtained from the eigenvectors of \autoref{eq:vmatrix}. The phases of the eigenvectors were adjusted to avoid discontinuities between different configurations in coordinate space.

\section{Setup of the assignment}
\label{sec:setup_assignment}
\begin{figure*}
\includegraphics[scale=.8]{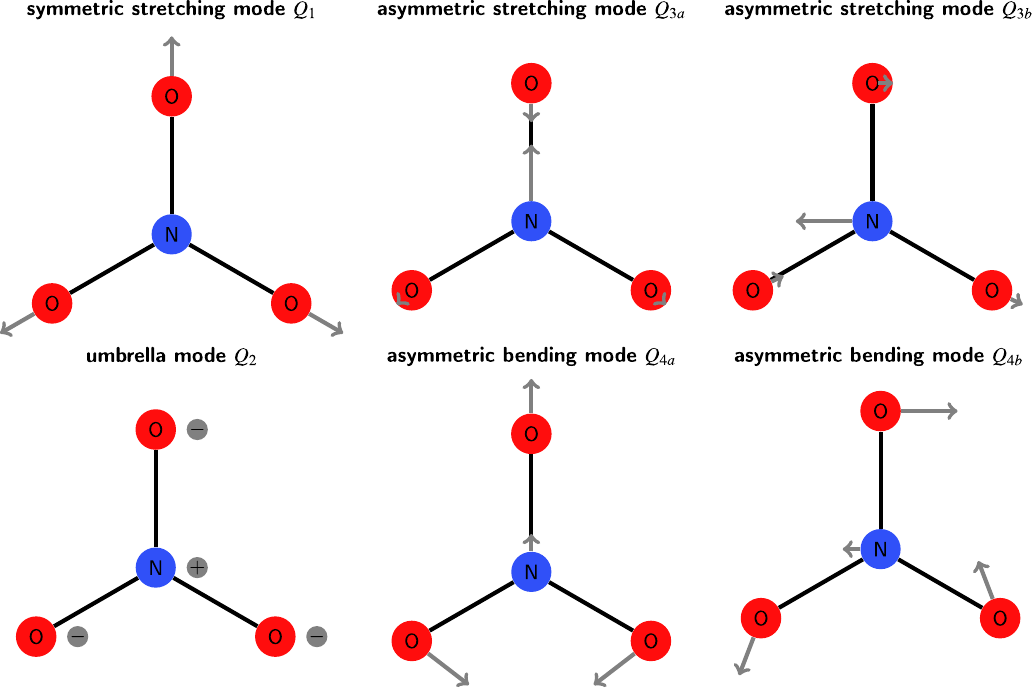}
  \caption{
  Normal modes used for the assignment. For each modes the gray arrows or circles with $\pm$ signs display the displacement vectors. The circles with $\pm$ signs  display motions out of plane. The displacement vectors (not the circles) are drawn to scale of the actually used modes, hence the small extent of some of the vectors.}
  \label{fig:modes}
\end{figure*}

We assigned the vibronic states not using curvilinear but normal modes.
They are displayed in \autoref{fig:modes}.
$Q_1$ corresponds to the totally symmetric stretch motion, $Q_2$ to the umbrella motion, the degenerate $Q_{3a}$ and $Q_{3b}$ to the antisymmetric stretch motions, and 
 $Q_{4a}$ and $Q_{4b}$ to the antisymmetric bending motions. The symmetry irreps in $D_{3h}$ are $a_1^{'}$, $a_2^{''}$, $e^{'}$, $e^{'}$, respectively.
 Note that these symmetry-adapted normal modes differ from textbook normal modes for \ce{AB3}-type molecules.\cite{Pictorial2006merlina}
 The convention we used to define the normal modes were such that the $Q_{3a}$ and $Q_{4a}$ modes are mirror-symmetric along the $xz$-plane, hence leading to $C_{2v}$-symmetric geometries, while the $Q_{3b}$ and $Q_{4b}$ lead to $C_s$-symmetric geometries.

We assigned the states
mostly based on wavefunction cuts, overlaps to vibrational self-consistent field (VSCF) states,
energies, and symmetry considerations.
We optimized 
the VSCF\cite{Self1978bowman} states
using normal modes, keeping the degenerate $Q_{3a}/Q_{3b}$ and the $Q_{4a}/Q_{4b}$ mode pairs correlated. We individually optimized excited VSCF states.

Plotting the two-dimensional wavefunction cuts requires deciding which coordinates are kept frozen.
Following \lit{Stateresolved2022larsson},
we determined the values of the frozen coordinates
for the two-dimensional wavefunction cuts
from the maxima of the diagonal of the one-dimensional reduced density matrices (reduced densities). In cases of several maxima of the densities, we generated all possible combinations.
We chose the cuts presented here to be most representative.

To generate the wavefunction cuts and the density matrices, we
transformed the wavefunction from the TTNS represented on the direct-product grid in curvilinear coordinates to a state represented in normal coordinates on a small direct-product grid.
Specifically, we converted the normal-mode grid-points to curvilinear coordinates and then computed the amplitudes of the states on these points by interpolating the DVR functions and contracting the TTNS. 
Finally, we converted the diabatic state represented on the small direct-product grid to the adiabatic representation in order to plot the vibrational wavefunction on the adiabatic ground state only.

We obtained the symmetries for the diabatic ground state only using the $C_{2v}$ point group and the symmetry relationships for the curvilinear coordinates described in \lit{NO32018viel}.  Given the $e'$ symmetry of the two $Q_3$ and $Q_4$ modes, we used symmetry considerations to derive the number of energies and their associated irreps for each specific quanta in $Q_3$ and $Q_4$.
\autoref{tab:c2vtod3h} lists the irreps in the $C_{2v}$ and $D_{3h}$ point groups and their relationships to the used normal modes.
Note that throughout we use the symmetries for the \emph{vibrational} state on the diabatic state. Other authors list the \emph{vibronic} symmetries that combines the vibrational symmetry and that of the electronic state. The relationship between vibrational and vibronic irreps is shown in the last two columns of \autoref{tab:c2vtod3h}.

\begin{table}
\caption{Relation between (``vibrational'') irreps in $C_{2v}$ and $D_{3h}$ within the orientational convention of this work similar to \lit{NO32018viel}.
The normal modes are listed in the line corresponding to their irrep in $C_{2v}$.
The last two columns provide the relations of the irreps when the symmetry of the lowest diabatic state ($A_2'$) is taken into account (``vibronic'' irreps).
}
\label{tab:c2vtod3h}
 \begin{tabular*}{.48\textwidth}{@{\extracolsep{\fill}}rrc|rr}
    \toprule
 $\Gamma_{C_{2v}}$ & $\Gamma_{D_{3h}}$ & normal mode &$\Gamma_{C_{2v}\cdot A_2'}$ & $\Gamma_{D_{3h}\cdot A_2'}$  \\ \midrule
 $a_1$ &  $a_1'$ or $e'$     &  $Q_1$, $Q_{3a}$, $Q_{4a}$       & $b_1$& $a_2'$ or $e'$    \\
 $b_1$ &  $a_2'$ or $e'$     &  $Q_{3b}$, $Q_{4b}$ & $a_1$& $a_1'$ or $e'$    \\
 $a_2$ &  $a_1''$ or $e''$   &          & $b_2$& $a_2''$ or $e''$  \\
 $b_2$ &  $a_2''$ or $e''$   &     $Q_2$& $a_2$& $a_1''$ or $e''$  \\
    \bottomrule
  \end{tabular*}
  \end{table}

\section{Results and discussion}
\label{sec:results}
\subsection{Overview}
\label{subsec:results_overview}

\begin{figure}
\includegraphics[scale=.9]{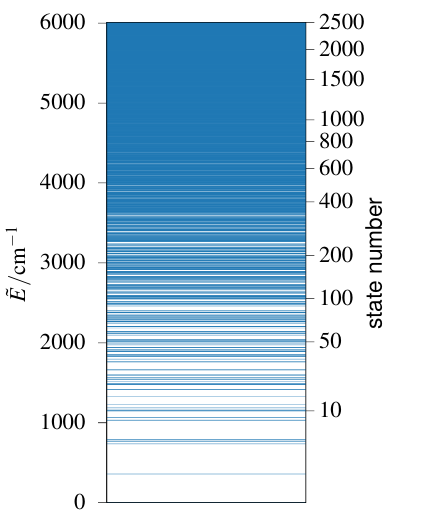}
  \caption{Computed energy levels. A line is plotted for each level. The left (right) ordinate displays the wavenumber (state number).
  The vibronic ground state energy is taken as zero point. }
  \label{fig:stick_spectrum}
\end{figure}

\begin{figure}
\includegraphics[scale=.9]{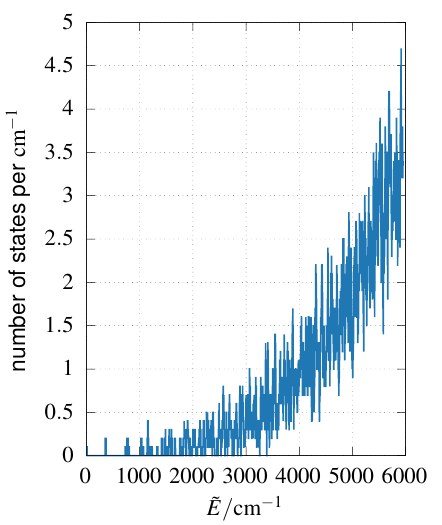}
  \caption{Density of states as a function of energy. For each term the number of states is shown within $\unit[10]{\icm}$, divided by $\unit[10]{\icm}$. Note that degenerate states are counted separately.}
  \label{fig:dos}
\end{figure}

Overall, we computed approximately 2500 eigenstates up to an excitation energy of  $\tilde E = \unit[6000]{\icm}$, and some more states up to $\tilde E \approx \unit[6365]{\icm}$.
In the following, all energies/wavenumbers stated are relative to the vibronic ground state energy of $\unit[2396]{\icm}$. 
\autoref{fig:stick_spectrum} gives an overview of the computed energy levels up to $\unit[6000]{\icm}$. 
Note that the electronic $\tilde A ^2 E''$ state appears at $\tilde E \approx \unit[7065]{\icm}$, so our high-energy vibronic states have energies just $\unit[700]\icm$ below the next electronic state.
At these high energies, there is a large density of states and an apparent quasi-continuum.\cite{Multimode1984koppel}
For comparison, there are 400 vibronic states up to $\unit[3761]{\icm}$, 1000 states up to $\unit[4787]{\icm}$, but already approximately 2500 states up to $\unit[6000]{\icm}$.
The approximate density of states as a function of excitation energy is shown in \autoref{fig:dos} and displays a steep increase of the number of states from $\sim 0.5$ states per $\unit{\icm}$ at $\unit[3000]{\icm}$ to almost $4$ states   per $\unit{\icm}$ at $\unit[6000]{\icm}$. 
Note that our algorithm does not necessarily return eigenstates ordered by energy. 
While we did additional computations to retrieve all  states (by designing highly excited VSCF-optimized initial guesses for the TTNS optimization), 
there might be some states above $\sim\unit[5800]{\icm}$ that we missed and that would lead to a slight change in the density of states at these high energies. 
Since most of the following analysis is for the states below $\unit[3000]{\icm}$, the missing states at these high energies are irrelevant. 

The bond dimensions used for re-fitting the PES as MPOs ($D=600$ and $D=1,800$) are extremely large for 6-dimensional PESs. Using weights in the least-square MPO fit can reduce the required basis size/bond dimension in related fitting methods.\cite{Product1996jackle} Here, weights did not reduce the required bond dimension. This might be explained by the 
importance of large-energy regions on one electronic state due to the large diabatic couplings as well as the fact that low-energy regions can have large couplings.
We also tried other fitting  methods such as ``potfit'' or canonical decompositions, which recently have been generalized to diabatic PESs,\cite{Representation2022han} but found that MPOs provide the most straightforward and accurate refits.
The $D=600$ MPO representation of the lowest diabatic PES has a root-mean-square deviation to the fitted PES (RMSD) of $\unit[21]{\icm}$ 
whereas the couplings and excited states have a maximal RMSD of $\unit[62]{\icm}$.
The RMSD for the $D=1,800$ MPO are $\unit[16]{\icm}$ for the ground state and  up to 
$\unit[25]{\icm}$ for the excited states and couplings.
Note that these RMSDs reflect the mean accuracy of the PES along the complete configuration space covered. Typically, the eigenstate observables are much more accurate than the RMSD of the PES.\cite{Stateresolved2022larsson}
However, degeneracies of high-energy states are better described by MPOs with larger bond dimension. 

To estimate the error of the MPO approximation of the potential energy operator, we have checked the energy error of the first 500 and the last 160 states using the DP-DVR method by converting our TTNS state to a DP-DVR state and computing the DP-DVR energy using the exact PES without MPO fit; see \autoref{sec:setup} for more details.
The DP-DVR energies differ mostly  by up to $\unit[2]{\icm}$ to the TTNS-MPO energy, with maximal errors reaching  $\sim\unit[3.0]{\icm}$ for the lowest 500 states and $\sim\unit[3.6]{\icm}$ for the highest 160 states. This shows that the error only modestly increases with excitation energy.
Note that the approximate DP-DVR energies are converged to $\precsim~\unit[2]{\icm}$, indicating that the actual error of our TTNS-MPO energies often are much smaller than $\unit[2]{\icm}$. 
This accuracy is more than sufficient and much higher than the accuracy of the underlying artificial-neural-network-based PES.

\subsection{Assignment}
\label{subsec:assignment}

\enlargethispage{1\baselineskip} %
The assignment in terms of vibrational zero-order states of the first 180 vibronic states up to $\tilde E \approx \unit[3000]{\icm}$ is shown in \autoref{tab:assignment},  
and their energies are compared with experiment and three other calculations from literature by Viel and Eisfeld,\cite{NO32018viel}  Stanton,\cite{Simulation2022stanton} and by Homayoon and Bowman\cite{Communication2014homayoon} (note that the latter is based on an empirical coordinate scaling).
In addition, \autoref{tab:assignment} contains symmetry labels for each vibronic level.
In particular, therein we provide a ``grouped assignment'' using Roman numerals to pinpoint the number of energies and the explicit list of irreps for each combination of quanta in $Q_3$ and $Q_4$.
For example, two quanta in $Q_4$ (abbreviated as $4^2$)  leads to two irreps $a_1'$ and $e'$, and hence two levels for the group we denote as ``$III: k/2 (a_1' + e')$,'' with $k=1,2$ numbering the two levels in the group.
The first level ($k=1$) in that group corresponds to state number 3 in \autoref{tab:assignment}, which is of $a_1$ symmetry, while the states 4/5 correspond to the second level ($k=2$) in group $III$ with $e'$ symmetry.
For each row in the table, the irrep of the particular state considered in the row is underlined.
The groups are also used for combination terms to denote excitations in both $Q_1$ or $Q_2$ and $Q_3$ or $Q_4$.
Note that,  due to the $a_2''$ symmetry of the $Q_2$ mode,
the irreps change if the group is combined with an odd excitation in $Q_2$.

\if\USETWOCOLUMNS1
\onecolumn
\fi
\begin{landscape}
\begingroup
\if\USETWOCOLUMNS1
\small
\else
\tiny
\fi
\begin{longtable}{@{\extracolsep{\fill}}rr%
  llll%
  ll%
  l%
  rllr%
  rr}
\caption{%
Assignment of the first 180 vibronic states up to $\sim\unit[3000]{\icm}$. 
For each level, we present the vibrational irrep of the $C_{2v}$ point group,
the quanta $v_i$ for mode $i$, a grouped assignment (see below), and the wavenumbers computed in this work, 
absolute difference to experimental wavenumber $\Delta$,
experimental wavenumbers (as assigned) and computed wavenumbers from the literature. 
The zero-point vibrational energy corresponds to $\unit[2396.32]{\icm}$.
Note that the vibrational irrep corresponds to the vibrational symmetry of the vibronic wavefunction component onto the lowest diabatic state of $A_2'$ symmetry; see \autoref{tab:c2vtod3h} for conversions to other point groups or vibronic irreps.
If explicit assignments can be made for the degenerate 
antisymmetric stretching and bending coordinates ($v_3$/$v_4$), then we will show it using Dirac notation. Otherwise, we only show the main vibrational quantum number. 
The grouped assignment provides an alternative labeling ($Q^n$ corresponds to $n$ quanta in mode $Q$) and a grouping of the excitations in $v_3$ and $v_4$ together with the vibrational irreps of that group using the $D_{3h}$ point group; see text for details.
The groups are numbered using Roman numerals.
Unclear assignments are marked by parenthesizing the grouped assignment. Some levels were too difficult to assign, which is marked by an asterisk. %
The experimental wavenumbers are shown with up to 6 digits precision. If the uncertainty was provided in the literature but actually is smaller than the precision reported here, then the  uncertainty is set to 0. 
``S'' in the experimental references refers to summation of low-energy levels as calculated in \lit{Infrared2017kawaguchi}.
Note that the two exp.~levels marked in parentheses from  \lit{Infrared2008jacox} are based on matrix isolation.
Note further that some experimental values, in particular some with large deviation to our levels, were not directly observed but deduced from perturbation analysis; see the text for a discussion.
Some computed levels in \lit{NO32018viel} were not assigned. These are marked with a ``?''.
States that may appear higher in energy are marked as ``missing''.
The computations reported in \lit{Simulation2022stanton} did not include the umbrella $Q_2$ mode, thus the ``n/a'' label for those states. Some levels using a different potential energy surface and a small basis that includes the $Q_2$ mode were reported by the same author in \lit{Vibronic2009stantona}. If available, these levels are added in parentheses after the ``n/a'' label.
Note that the levels reported in \lit{Communication2014homayoon}  are based on a semi-empirical coordinate scaling and thus are not from \textit{ab initio} computations.
  } 
  \label{tab:assignment}\\
    \toprule
   & & & & & & & &  \multicolumn{7}{c}{$\tilde E/\unit{\icm}$}\\
    State &  $\Gamma_{C_{2v}}$ &%
    $v_1$ & $v_2$ & $v_3$ & $v_4$ & %
    \multicolumn{2}{l}{grouped assignment} & %
   this work & $\Delta$ & Exp. & Ref. & \lit{NO32018viel} & \lit{Simulation2022stanton} &    \lit{Communication2014homayoon}  %
    \\
    \midrule
  \endfirsthead
   & & & & & & & &  \multicolumn{7}{c}{$\tilde E/\unit{\icm}$}\\
    State &  $\Gamma_{C_{2v}}$ &%
    $v_1$ & $v_2$ & $v_3$ & $v_4$ & %
    \multicolumn{2}{l}{grouped assignment} & %
   this work & $\Delta$ & Exp. & Ref. & \lit{NO32018viel} & \lit{Simulation2022stanton} &    \lit{Communication2014homayoon}  %
    \\
    \midrule
    \endhead
0 & $a_1$ &  &  &  &  &  & I: $1/1$ ($\underline{a_1'}$) & $0$\\
1/2 & $a_1/b_1$ &  &  &  & $\ket{01}/\ket{10}$ & $4^1$ & II: $1/1$ ($\underline{e'}$) & $359.33$ & $6$ & 365.49(0) & [\!\!\citenum{Fourier2013fujimori}]  & 361.1 & 369 & 369\\
3 & $a_1$ &  &  &  & $\ket{02}+\ket{20}$ & $4^2$ & III: $1/2$ ($\underline{a_1'}+e'$) & $736.09$ & $16$ & 752.40(0) & [\!\!\citenum{Fourier2013fujimori,Infrared2017kawaguchi}]  & 711 & 741 & 746\\
4/5 & $a_1/b_1$ &  &  &  & $\ket{11}/\ket{02}-\ket{20}$ & $4^2$ & III: $2/2$ ($a_1'+\underline{e'}$) & $770.68$ & $1$ & 771.79(1) & [\!\!\citenum{Fourier2013fujimori,Infrared2017kawaguchi}]  & 742.2 & 777 & 756\\
6 & $b_2$ &  & $1$ &  &  &$2^1$ & &  $792.73$ & $30$ & 762.34(0) & [\!\!\citenum{Fourier2013fujimori,Infrared2017kawaguchi}]  & 748.5 & n/a (808) & 764\\
7/8 & $a_1/b_1$ &  &  & $\ket{01}/\ket{10}$ &  & $3^1$ & IV: $1/1$ ($\underline{e'}$) & $1029.01$ & $25$ & 1054.14(0) & [\!\!\citenum{Infrared2021kawaguchi}]  & 1021.8 & 1069 & 1099\\
9 & $a_1$ & $1$ &  &  &  &$1^1$ & &  $1064.45$ & $13$ & 1051.26 & [\!\!\citenum{Laser2022fukushima}]  & 1038.6 & 1061 & 1067\\
10/11 & $a_2/b_2$ &  & $1$ &  & $\ket{01}/\ket{10}$ & $2^{1} \cdot4^1$ & $2^{1} \cdot$II: $1/1$ ($\underline{e''}$) & $1146.88$ & $22$ & 1125.10(0) & [\!\!\citenum{Infrared2017kawaguchi}]  & 1109.6 & n/a & 1138\\
12/13 & $a_1/b_1$ &  &  &  & $3$ & $4^3$ & V: $1/3$ ($a_1'+a_2'+\underline{e'}$) & $1158.68$ & $15$ & 1173.61(0) & [\!\!\citenum{Infrared2018kawaguchia}]  & 1082.5 & 1152 & 1143\\
14 & $b_1$ &  &  &  & $\ket{21}+\ket{03}$ & $4^3$ & V: $2/3$ ($a_1'+\underline{a_2'}+e'$) & $1185.08$ & $31$ & 1216.2 & [\!\!\citenum{Laser2022fukushima}]  & 1134.6 & 1214 & 1150\\
15 & $a_1$ &  &  &  & $3$ & $4^3$ & V: $3/3$ ($\underline{a_1'}+a_2'+e'$) & $1224.77$ & $169$ & 1055.34 & [\!\!\citenum{Laser2022fukushima}]  & 1139.7 & 1191 & 1160\\
16 & $b_1$ &  &  & $1$ & $1$ & $3^14^1$ & VI: $1/3$ ($a_1'+\underline{a_2'}+e'$) & $1329.90$ & $161$ & 1491(3) & [\!\!\citenum{FTIR2013kawaguchi,Infrared2017kawaguchi}]  & 1302.4 & 1365 & 1430\\
17/18 & $a_1/b_1$ & $1$ &  &  & $\ket{01}/\ket{10}$ & $1^{1} \cdot4^1$ & $1^{1} \cdot$II: $1/1$ ($\underline{e'}$) & $1418.46$ & $5$ & 1413.57(0) & [\!\!\citenum{Infrared2018kawaguchia}]  & 1388.1 & 1424 & 1434\\
19/20 & $a_1/b_1$ &  &  & $1$ & $1$ & $3^14^1$ & VI: $2/3$ ($a_1'+a_2'+\underline{e'}$) & $1476.71$ & $16$ & 1492.39(0) & [\!\!\citenum{FTIR2013kawaguchi,Infrared2017kawaguchi}]  & 1438.6 & 1494 & 1493\\
21 & $a_1$ &  &  & $1$ & $1$ & $3^14^1$ & VI: $3/3$ ($\underline{a_1'}+a_2'+e'$) & $1490.45$ & $9$ & 1499.75(0) & [\!\!\citenum{FTIR2013kawaguchi,Infrared2017kawaguchi}]  & 1425.0 & 1522 & 1501\\
22 & $b_2$ &  & $1$ &  & $\ket{02}+\ket{20}$ & $2^{1} \cdot4^2$ & $2^{1} \cdot$III: $1/2$ ($\underline{a_2''}+e''$) & $1516.41$ & $7$ & 1509.72(2) & [\!\!\citenum{FTIR2013kawaguchi,Infrared2017kawaguchi}]  & 1459.2 & n/a (1598) & 1519\\
23 & $a_1$ &  & $2$ &  &  &$2^2$ & &  $1542.69$ & $21$ & 1522(2) & [\!\!\citenum{FTIR2013kawaguchi,Infrared2017kawaguchi}]  & 1496.2 & n/a & 1533\\
24/25 & $a_2/b_2$ &  & $1$ &  & $\ket{11}/\ket{02}-\ket{20}$ & $2^{1} \cdot4^2$ & $2^{1} \cdot$III: $2/2$ ($a_2''+\underline{e''}$) & $1559.70$ & $22$ & 1537.54(2) & [\!\!\citenum{FTIR2013kawaguchi,Infrared2017kawaguchi}]  & 1490.2 & n/a & 1529\\
26 & $a_1$ &  &  &  & $4$ & $4^4$ & VII: $1/3$ ($\underline{a_1'}+e'+e'$) & $1587.41$ & $22$ & 1609(10) & [\!\!\citenum{Infrared2017kawaguchi}]  & 1446.9 & 1569 & 1553\\
27/28 & $a_1/b_1$ &  &  &  & $\ket{04}/\ket{31}$ & $4^4$ & VII: $2/3$ ($a_1'+\underline{e'}+e'$) & $1597.92$ & $31$ & 1567 & [\!\!\citenum{HighResolution2020babin}]  & 1469.1 & 1579 & 1552\\
29/30 & $a_1/b_1$ &  &  &  & $4$ & $4^4$ & VII: $3/3$ ($a_1'+e'+\underline{e'}$) & $1661.64$ &  &  &   & 1542.0 & 1642 & 1568\\
31/32 & $a_1/b_1$ &  &  & $\ket{01}/\ket{10}$ & $\ket{11}/\ket{20}$ & $3^14^2$ & VIII: $1/4$ ($a_1'+a_2'+\underline{e'}+e'$) & $1763.73$ & $186$ & 1949.83(1) & [\!\!\citenum{Infrared2017kawaguchi}]  & 1678.2 & 1769 & 1804\\
33 & $a_1$ & $1$ &  &  & $\ket{20}+\ket{02}$ & $1^{1} \cdot4^2$ & $1^{1} \cdot$III: $1/2$ ($\underline{a_1'}+e'$) & $1796.97$ & $2$ & 1798.5 & [\!\!\citenum{Laser2022fukushima}]  & 1736.4 & 1792 & 1805\\
34/35 & $a_2/b_2$ &  & $1$ & $\ket{01}/\ket{10}$ &  & $2^{1} \cdot3^1$ & $2^{1} \cdot$IV: $1/1$ ($\underline{e''}$) & $1823.72$ & $9$ & 1815 & S  & 1763.1 & n/a & 1858\\
36/37 & $a_1/b_1$ & $1$ &  &  & $\ket{11}$ & $1^{1} \cdot4^2$ & $1^{1} \cdot$III: $2/2$ ($a_1'+\underline{e'}$) & $1844.97$ & $11$ & 1834.3 & [\!\!\citenum{Laser2022fukushima}]  & 1772.6 & 1845 & 1832\\
38 & $b_2$ & $1$ & $1$ &  &  &$1^1 2^1$ & &  $1855.21$ & $45$ & 1810 & S  & 1783.5 & n/a (1898) & 1830\\
39/40 & $a_1/b_1$ &  & $2$ &  & $\ket{01}/\ket{10}$ & $2^2 \cdot4^1$ & $2^2 \cdot$II: $1/1$ ($\underline{e'}$) & $1891.93$ & $7$ & 1885 & S  & 1857.3 & n/a & missing\\
41 & $b_1$ &  &  & $1$ & $\ket{12}$ & $3^14^2$ & VIII: $2/4$ ($a_1'+\underline{a_2'}+e'+e'$) & $1897.39$ & $17$ & 1914.18(0) & [\!\!\citenum{Infrared2017kawaguchi}]  & 1803.9 & 1889 & 1890\\
42 & $a_1$ &  &  & $1$ & $2$ & $3^14^2$ & VIII: $3/4$ ($\underline{a_1'}+a_2'+e'+e'$) & $1906.74$ & $14$ & 1920.76(0) & [\!\!\citenum{Infrared2017kawaguchi}]  & 1861.3? & 1923 & 1897\\
43/44 & $a_1/b_1$ &  &  & $1$ & $2$ & $3^14^2$ & VIII: $4/4$ ($a_1'+a_2'+e'+\underline{e'}$) & $1915.40$ & $11$ & 1926.15(0) & [\!\!\citenum{Infrared2017kawaguchi}]  & 1810.7 & 1931 & 1900\\
45/46 & $a_2/b_2$ &  & $1$ &  & $3$ & $2^{1} \cdot4^3$ & $2^{1} \cdot$V: $1/3$ ($a_1''+a_2''+\underline{e''}$) & $1939.03$ & $10$ & 1929.37(0) & [\!\!\citenum{Infrared2017kawaguchi}]  & 1830.1 & n/a & \\
47 & $a_2$ &  & $1$ &  & $\ket{21}+\ket{03}$ & $2^{1} \cdot4^3$ & $2^{1} \cdot$V: $2/3$ ($\underline{a_1''}+a_2''+e''$) & $1972.90$ & $35$ & 1937.68(8) & [\!\!\citenum{Infrared2017kawaguchi}]  & missing & n/a & \\
48 & $a_1$ &  &  & $\ket{20}+\ket{02}$ &  & $3^2$ & X: $1/2$ ($\underline{a_1'}+e'$) & $1995.12$ & $14$ & 2009.12(0) & [\!\!\citenum{Infrared2022kawaguchi}]  & 1881.7? & missing & \\
49/50 & $a_1/b_1$ & $1$ &  & $1$ &  & $1^{1} \cdot3^1$ & $1^{1} \cdot$IV: $1/1$ ($\underline{e'}$) & $2009.15$ & $15$ & 2024.32(0) & [\!\!\citenum{Infrared2018kawaguchi}]  & 1831.2? & missing & \\
51 & $b_2$ &  & $1$ &  & $3$ & $2^{1} \cdot4^3$ & $2^{1} \cdot$V: $3/3$ ($a_1''+\underline{a_2''}+e''$) & $2020.94$ & $51$ & 1970(10) & [\!\!\citenum{Infrared2017kawaguchi}]  & 1882.0? & n/a & \\
52/53 & $a_1/b_1$ &  &  &  & $\ket{05}/\ket{50}$ & $4^5$ & IX: $1/3$ ($a_1'+\underline{e'}+e'$) & $2031.72$ & $5$ & 2026.3 & [\!\!\citenum{Laser2022fukushima}]  & missing & 1989 & \\
54 & $b_1$ &  &  &  & $6$ & $4^6$ & XI: $1/5$ ($a_1'+a_1'+\underline{a_2'}+e'+e'$) & $2041.77$ &  &  &   & 1860.4? &  & \\
55 & $a_1$ &  &  &  & $5$ & $4^5$ & IX: $2/3$ ($\underline{a_1'}+e'+e'$) & $2095.81$ & $86$ & 2010.0 & [\!\!\citenum{Laser2022fukushima}]  & 1914.2? &  & \\
56/57 & $a_1/b_1$ &  &  &  & $5$ & $4^5$ & IX: $3/3$ ($a_1'+e'+\underline{e'}$) & $2115.12$ &  &  &   & missing &  & \\
58 & $a_2$ &  & $1$ & $1$ & $1$ & $2^{1} \cdot3^14^1$ & $2^{1} \cdot$VI: $1/3$ ($\underline{a_1''}+a_2''+e''$) & $2123.01$ &  &  &   & 1887.0? & n/a (2248) & \\
59 & $a_1$ & $2$ &  &  &  &$1^2$ & &  $2138.35$ & $21$ & 2117.8 & [\!\!\citenum{Laser2022fukushima}]  &  &  & \\
60/61 & $a_1/b_1$ &  &  & $2$ &  & $3^2$ & X: $2/2$ ($a_1'+\underline{e'}$) & $2139.84$ & $15$ & 2155.19(0) & [\!\!\citenum{Infrared2018kawaguchi}]  &  &  & \\
62 & $b_1$ &  &  & $1$ & $3$ & $3^14^3$ & XII: $1/5$ ($a_1'+\underline{a_2'}+e'+e'+e'$) & $2186.13$ &  &  &   &  &  & \\
63/64 & $a_1/b_1$ & $1$ &  &  & $3$ & $1^{1} \cdot4^3$ & $1^{1} \cdot$V: $1/3$ ($a_1'+a_2'+\underline{e'}$) & $2201.43$ & $4$ & 2205.69(0) & [\!\!\citenum{Infrared2018kawaguchi}]  &  &  & \\
65/66 & $a_2/b_2$ & $1$ & $1$ &  & $\ket{01}/\ket{10}$ & $1^1 2^1 \cdot4^1$ & $1^1 2^1 \cdot$II: $1/1$ ($\underline{e''}$) & $2205.46$ & $34$ & 2171.1 & [\!\!\citenum{Infrared2022kawaguchi}]  &  &  & \\
67/68 & $a_1/b_1$ &  &  & $1$ & $3$ & $3^14^3$ & XII: $2/5$ ($a_1'+a_2'+\underline{e'}+e'+e'$) & $2245.81$ &  &  &   &  &  & \\
69 & $a_1$ &  & $2$ &  & $\ket{02}+\ket{20}$ & $2^2 \cdot4^2$ & $2^2 \cdot$III: $1/2$ ($\underline{a_1'}+e'$) & $2263.33$ & $17$ & 2245.87(0) & [\!\!\citenum{Infrared2018kawaguchi}]  &  &  & \\
70/71 & $a_2/b_2$ &  & $1$ & $1$ & $\ket{11}/\ket{02}-\ket{20}$ & $2^{1} \cdot3^14^1$ & $2^{1} \cdot$VI: $2/3$ ($a_1''+a_2''+\underline{e''}$) & $2263.52$ &  &  &   &  &  & \\
72 & $b_1$ & $1$ &  &  & $\ket{21}+\ket{03}$ & $1^{1} \cdot4^3$ & $1^{1} \cdot$V: $2/3$ ($a_1'+\underline{a_2'}+e'$) & $2269.55$ &  &  &   &  &  & \\
73 & $b_2$ &  & $1$ & $1$ & $1$ & $2^{1} \cdot3^14^1$ & $2^{1} \cdot$VI: $3/3$ ($a_1''+\underline{a_2''}+e''$) & $2282.60$ &  &  &   &  &  & \\
74 & $a_1$ & $1$ &  &  & $3$ & $1^{1} \cdot4^3$ & $1^{1} \cdot$V: $3/3$ ($\underline{a_1'}+a_2'+e'$) & $2292.61$ & $135$ & 2157.8 & [\!\!\citenum{Laser2022fukushima}]  &  &  & \\
75/76 & $a_1/b_1$ &  & $2$ &  & $\ket{11}/\ket{02}-\ket{20}$ & $2^2 \cdot4^2$ & $2^2 \cdot$III: $2/2$ ($a_1'+\underline{e'}$) & $2297.43$ & $54$ & 2351.4 & [\!\!\citenum{Laser2022fukushima}]  &  &  & \\
77 & $b_2$ &  & $3$ &  &  &$2^3$ & &  $2315.93$ & $66$ & 2249.9 & [\!\!\citenum{Infrared2022kawaguchi}]  &  &  & \\
78/79 & $a_1/b_1$ &  &  & $1$ & $3$ & $3^14^3$ & XII: $3/5$ ($a_1'+a_2'+e'+\underline{e'}+e'$) & $2330.42$ &  &  &   &  &  & \\
80 & $b_1$ & $1$ &  & $1$ & $1$ & $1^{1} \cdot3^14^1$ & $1^{1} \cdot$VI: $1/3$ ($a_1'+\underline{a_2'}+e'$) & $2343.23$ &  &  &   &  &  & \\
81 & $a_1$ &  &  & $1$ & $3$ & $3^14^3$ & XII: $4/5$ ($\underline{a_1'}+a_2'+e'+e'+e'$) & $2349.16$ & $9$ & 2358 & [\!\!\citenum{NO31992kim}]  &  &  & \\
82 & $b_2$ &  & $1$ &  & $4$ & $2^{1} \cdot4^4$ & $2^{1} \cdot$VII: $1/3$ ($\underline{a_2''}+e''+e''$) & $2370.62$ &  &  &   &  &  & \\
83/84 & $a_1/b_1$ &  &  & $1$ & $3$ & $3^14^3$ & XII: $5/5$ ($a_1'+a_2'+e'+e'+\underline{e'}$) & $2376.78$ & $0$ & 2376.52(0) & [\!\!\citenum{Infrared2018kawaguchi}]  &  &  & \\
85/86 & $a_2/b_2$ &  & $1$ &  & $4$ & $2^{1} \cdot4^4$ & $2^{1} \cdot$VII: $2/3$ ($a_2''+\underline{e''}+e''$) & $2379.11$ & $19$ & 2360.0 & [\!\!\citenum{Infrared2022kawaguchi}]  &  &  & \\
87/88 & $a_1/b_1$ &  &  & $2$ & $1$ & $3^24^1$ & XIV: $1/4$ ($\underline{e'}+a_2'+a_1'+e'$) & $2399.18$ &  &  &   &  &  & \\
89/90 & $a_2/b_2$ &  & $1$ &  & $4$ & $2^{1} \cdot4^4$ & $2^{1} \cdot$VII: $3/3$ ($a_2''+e''+\underline{e''}$) & $2459.11$ &  &  &   &  &  & \\
91 & $a_1$ &  &  &  & $6$ & $4^6$ & XI: $2/5$ ($\underline{a_1'}+a_1'+a_2'+e'+e'$) & $2481.49$ &  &  &   &  &  & \\
92/93 & $a_1/b_1$ &  &  &  & $6$ & $4^6$ & XI: $3/5$ ($a_1'+a_1'+a_2'+\underline{e'}+e'$) & $2485.13$ &  &  &   &  &  & \\
94/95 & $a_1/b_1$ & $2$ &  &  & $\ket{01}/\ket{10}$ & $1^2 \cdot4^1$ & $1^2 \cdot$II: $1/1$ ($\underline{e'}$) & $2496.08$ &  &  &   &  &  & \\
96 & $a_1$ &  &  & $2$ & $1$ &($3^24^1$ & XIV: $2/4$ ($e'+a_2'+\underline{a_1'}+e'$) )& $2501.21$ &  &  &   &  &  & \\
97/98 & $a_1/b_1$ & $1$ &  & $1$ & $1$ & $1^{1} \cdot3^14^1$ & $1^{1} \cdot$VI: $2/3$ ($a_1'+a_2'+\underline{e'}$) & $2513.31$ & $5$ & 2518.71(0) & [\!\!\citenum{Infrared2018kawaguchi}]  &  &  & \\
99 & $a_1$ & $1$ &  & $1$ & $1$ & $1^{1} \cdot3^14^1$ & $1^{1} \cdot$VI: $3/3$ ($\underline{a_1'}+a_2'+e'$) & $2542.83$ & $32$ & 2511 & [\!\!\citenum{NO31992kim}]  &  &  & \\
100/101 & $a_2/b_2$ &  & $1$ & $1$ & $\ket{02}/\ket{20}$ & $2^{1} \cdot3^14^2$ & $2^{1} \cdot$VIII: $1/4$ ($a_1''+a_2''+\underline{e''}+e''$) & $2553.91$ &  &  &   &  &  & \\
102/103 & $a_1/b_1$ &  &  &  & $6$ & $4^6$ & XI: $4/5$ ($a_1'+a_1'+a_2'+e'+\underline{e'}$) & $2559.20$ &  &  &   &  &  & \\
104/105 & $a_1/b_1$ &  & $2$ & $\ket{01}/\ket{10}$ &  & $2^2 \cdot3^1$ & $2^2 \cdot$IV: $1/1$ ($\underline{e'}$) & $2569.23$ &  &  &   &  &  & \\
106 & $b_1$ &  &  &  & $7$ & $4^7$ & XVII: $1/5$ ($a_1'+\underline{a_2'}+e'+e'+e'$) & $2570.17$ &  &  &   &  &  & \\
107 & $b_2$ & $1$ & $1$ &  & $\ket{02}+\ket{20}$ & $1^1 2^1 \cdot4^2$ & $1^1 2^1 \cdot$III: $1/2$ ($\underline{a_2''}+e''$) & $2576.54$ & $69$ & 2508.0 & [\!\!\citenum{Infrared2022kawaguchi}]  &  &  & \\
108/109 & $a_1/b_1$ &  &  & $2$ & $1$ & $3^24^1$ & XIV: $3/4$ ($e'+a_2'+a_1'+\underline{e'}$) & $2577.33$ & $8$ & 2585.11(0) & [\!\!\citenum{Infrared2018kawaguchi}]  &  &  & \\
110 & $a_1$ &  &  &  & $6$ & $4^6$ & XI: $5/5$ ($a_1'+\underline{a_1'}+a_2'+e'+e'$) & $2590.42$ &  &  &   &  &  & \\
111 & $a_1$ & $1$ & $2$ &  &  &$1^1 2^2$ & &  $2592.77$ & $56$ & 2537.0 & [\!\!\citenum{Laser2022fukushima}]  &  &  & \\
112 & $b_1$ &  &  & $2$ & $\ket{11}$ &($3^24^1$ & XIV: $4/4$ ($e'+\underline{a_2'}+a_1'+e'$) )& $2615.17$ &  &  &   &  &  & \\
113/114 & $a_1/b_1$ &  &  & $1$ & $4$ & $3^14^4$ & XVI: $1/7$ ($a_1'+a_1'+a_2'+a_2'+\underline{e'}+e'+e'$) & $2622.58$ &  &  &   &  &  & \\
115/116 & $a_2/b_2$ & $1$ & $1$ &  & $\ket{11}/\ket{02}-\ket{20}$ & $1^1 2^1 \cdot4^2$ & $1^1 2^1 \cdot$III: $2/2$ ($a_2''+\underline{e''}$) & $2633.03$ &  &  &   &  &  & \\
117 & $a_1$ & $1$ &  &  & $4$ & $1^{1} \cdot4^4$ & $1^{1} \cdot$VII: $1/3$ ($\underline{a_1'}+e'+e'$) & $2639.21$ & $126$ & 2513.7 & [\!\!\citenum{Laser2022fukushima}]  &  &  & \\
118/119 & $a_2/b_2$ &  & $3$ &  & $\ket{01}/\ket{10}$ & $2^3 \cdot4^1$ & $2^3 \cdot$II: $1/1$ ($\underline{e''}$) & $2661.50$ &  &  &   &  &  & \\
120 & $a_2$ &  & $1$ & $1$ & $\ket{21}$ & $2^{1} \cdot3^14^2$ & $2^{1} \cdot$VIII: $2/4$ ($\underline{a_1''}+a_2''+e''+e''$) & $2677.76$ &  &  &   &  &  & \\
121/122 & $a_1/b_1$ &  & $2$ &  & $3$ & $2^2 \cdot4^3$ & $2^2 \cdot$V: $1/3$ ($a_1'+a_2'+\underline{e'}$) & $2683.08$ &  &  &   &  &  & \\
123/124 & $a_1/b_1$ &  &  & $1$ & $4$ & $3^14^4$ & XVI: $2/7$ ($a_1'+a_1'+a_2'+a_2'+e'+\underline{e'}+e'$) & $2686.85$ &  &  &   &  &  & \\
125 & $b_1$ &  &  & $1$ & $4$ & $3^14^4$ & XVI: $3/7$ ($a_1'+a_1'+\underline{a_2'}+a_2'+e'+e'+e'$) & $2693.42$ &  &  &   &  &  & \\
126 & $b_2$ &  & $1$ & $1$ & $3$ & $2^{1} \cdot3^14^2$ & $2^{1} \cdot$VIII: $3/4$ ($a_1''+\underline{a_2''}+e''+e''$) & $2694.96$ &  &  &   &  &  & \\
127/128 & $a_2/b_2$ &  & $1$ & $1$ & $2$ & $2^{1} \cdot3^14^2$ & $2^{1} \cdot$VIII: $4/4$ ($a_1''+a_2''+e''+\underline{e''}$) & $2705.88$ &  &  &   &  &  & \\
129 & $b_1$ &  & $2$ &  & $\ket{21}+\ket{03}$ & $2^2 \cdot4^3$ & $2^2 \cdot$V: $2/3$ ($a_1'+\underline{a_2'}+e'$) & $2713.70$ &  &  &   &  &  & \\
130 & $a_1$ &  &  & $1$ & $4$ & $3^14^4$ & XVI: $4/7$ ($\underline{a_1'}+a_1'+a_2'+a_2'+e'+e'+e'$) & $2717.39$ &  &  &   &  &  & \\
131/132 & $a_1/b_1$ & $1$ &  &  & $4$ & $1^{1} \cdot4^4$ & $1^{1} \cdot$VII: $2/3$ ($a_1'+\underline{e'}+e'$) & $2729.40$ &  &  &   &  &  & \\
133 & $a_1$ &  & $2$ &  & $3$ & $2^2 \cdot4^3$ & $2^2 \cdot$V: $3/3$ ($\underline{a_1'}+a_2'+e'$) & $2753.55$ & ($133$) & (2621) & [\!\!\citenum{Infrared2008jacox}]  &  &  & \\
134/135 & $a_1/b_1$ &  &  & $1$ & $4$ & $3^14^4$ & XVI: $5/7$ ($a_1'+a_1'+a_2'+a_2'+e'+e'+\underline{e'}$) & $2761.88$ &  &  &   &  &  & \\
136 & $b_1$ &  &  & $1$ & $4$ & $3^14^4$ & XVI: $6/7$ ($a_1'+a_1'+a_2'+\underline{a_2'}+e'+e'+e'$) & $2782.03$ &  &  &   &  &  & \\
137 & $b_2$ &  & $1$ & $2$ &  & $2^{1} \cdot3^2$ & $2^{1} \cdot$X: $1/2$ ($\underline{a_2''}+e''$) & $2786.72$ &  &  &   &  &  & \\
138 & $b_1$ &  &  & $1$ & * &$*$ & &  $2787.76$ &  &  &   &  &  & \\
139 & $a_1$ &  &  & $1$ & * &$*$ & &  $2787.96$ &  &  &   &  &  & \\
140/141 & $a_2/b_2$ & $1$ & $1$ & $1$ &  & $1^1 2^1 \cdot3^1$ & $1^1 2^1 \cdot$IV: $1/1$ ($\underline{e''}$) & $2805.05$ &  &  &   &  &  & \\
142 & $a_1$ &  &  & $2$ & $2$ &($3^24^2$ & XVIII: $1/6$ ($\underline{a_1'}+a_1'+a_2'+e'+e'+e'$) )& $2811.34$ &  &  &   &  &  & \\
143/144 & $a_2/b_2$ &  & $1$ &  & $5$ & $2^{1} \cdot4^5$ & $2^{1} \cdot$IX: $1/3$ ($a_2''+\underline{e''}+e''$) & $2814.45$ &  &  &   &  &  & \\
145 & $a_2$ &  & $1$ &  & $6$ & $2^{1} \cdot4^6$ & $2^{1} \cdot$XI: $1/5$ ($\underline{a_1''}+a_2''+a_2''+e''+e''$) & $2822.52$ &  &  &   &  &  & \\
146/147 & $a_1/b_1$ &  &  & $2$ & $2$ &($3^24^2$ & XVIII: $2/6$ ($a_1'+a_1'+a_2'+\underline{e'}+e'+e'$) )& $2826.90$ &  &  &   &  &  & \\
148 & $a_1$ &  &  & $1$ & $4$ & $3^14^4$ & XVI: $3/7$ ($a_1'+\underline{a_1'}+a_2'+a_2'+e'+e'+e'$) & $2852.13$ &  &  &   &  &  & \\
149 & $b_1$ &  & $2$ & $1$ & $1$ & $2^2 \cdot3^14^1$ & $2^2 \cdot$VI: $1/3$ ($a_1'+\underline{a_2'}+e'$) & $2865.51$ &  &  &   &  &  & \\
150/151 & $a_1/b_1$ &  &  & $2$ & $2$ &($3^24^2$ & XVIII: $3/6$ ($a_1'+a_1'+a_2'+e'+\underline{e'}+e'$) )& $2874.44$ & ($18$) & (2892) & [\!\!\citenum{Infrared2008jacox}]  &  &  & \\
152 & $a_1$ & $1$ &  & $2$ & $2$ &($1^{1} \cdot3^14^2$ & $1^{1} \cdot$VIII: $1/4$ ($\underline{a_1'}+a_2'+e'+e'$) )& $2880.45$ &  &  &   &  &  & \\
153/154 & $a_1/b_1$ &  &  & $3$ &  & $3^3$ & XIII: $1/3$ ($\underline{e'}+a_2'+a_1'$) & $2883.33$ &  &  &   &  &  & \\
155 & $b_2$ &  & $1$ &  & $5$ & $2^{1} \cdot4^5$ & $2^{1} \cdot$IX: $2/3$ ($\underline{a_2''}+e''+e''$) & $2886.64$ &  &  &   &  &  & \\
156 & $a_1$ & $2$ &  & $2$ & $2$ &($3^24^2$ & XVIII: $4/6$ ($a_1'+\underline{a_1'}+a_2'+e'+e'+e'$) )& $2892.68$ &  &  &   &  &  & \\
157 & $b_1$ &  &  & $3$ & $1$ &($3^34^1$ & XV: $1/6$ ($e'+e'+e'+\underline{a_2'}+a_1'+e'$) )& $2894.44$ &  &  &   &  &  & \\
158/159 & $a_2/b_2$ &  & $1$ &  & $5$ & $2^{1} \cdot4^5$ & $2^{1} \cdot$IX: $3/3$ ($a_2''+e''+\underline{e''}$) & $2915.59$ &  &  &   &  &  & \\
160/161 & $a_1/b_1$ & $2$ &  &  & $\ket{11}/\ket{02}-\ket{20}$ & $1^2 \cdot4^2$ & $1^2 \cdot$III: $1/2$ ($a_1'+\underline{e'}$) & $2916.70$ &  &  &   &  &  & \\
162 & $b_2$ & $2$ & $1$ &  &  &$1^2 2$ & &  $2925.70$ &  &  &   &  &  & \\
163/164 & $a_2/b_2$ &  & $1$ & $2$ &  & $2^{1} \cdot3^2$ & $2^{1} \cdot$X: $2/2$ ($a_2''+\underline{e''}$) & $2931.09$ &  &  &   &  &  & \\
165/166 & $a_1/b_1$ & $1$ & $2$ &  & $\ket{01}/\ket{10}$ & $1^1 2^2 \cdot4^1$ & $1^1 2^2 \cdot$II: $1/1$ ($\underline{e'}$) & $2939.55$ &  &  &   &  &  & \\
167/168 & $a_1/b_1$ &  &  &  & $7$ & $4^7$ & XVII: $2/5$ ($a_1'+a_2'+\underline{e'}+e'+e'$) & $2940.87$ &  &  &   &  &  & \\
169 & $b_1$ &  &  &  & $8$ & $4^8$ & XIX: $1/6$ ($a_1'+a_1'+\underline{a_2'}+e'+e'+e'$) & $2945.96$ &  &  &   &  &  & \\
170 & $b_1$ &  &  & $2$ & $2$ &($3^24^2$ & XVIII: $5/6$ ($a_1'+a_1'+\underline{a_2'}+e'+e'+e'$) )& $2956.48$ &  &  &   &  &  & \\
171/172 & $a_1/b_1$ & $1$ &  & $1$ & $2$ & $1^{1} \cdot3^14^2$ & $1^{1} \cdot$VIII: $2/4$ ($a_1'+a_2'+\underline{e'}+e'$) & $2966.44$ & $64$ & 2901.95(0) & [\!\!\citenum{Infrared2018kawaguchi}]  &  &  & \\
173 & $a_2$ &  & $1$ & $1$ & $3$ & $2^{1} \cdot3^14^3$ & $2^{1} \cdot$XII: $1/5$ ($\underline{a_1''}+a_2''+e''+e''+e''$) & $2975.60$ &  &  &   &  &  & \\
174/175 & $a_2$ & $1$ & $1$ &  & $3$ & $1^1 2^1 \cdot4^3$ & $1^1 2^1 \cdot$V: $1/3$ ($\underline{a_1''}+a_2''+e''$) & $2988.55$ &  &  &   &  &  & \\
176/177 & $a_1/b_1$ &  & $1$ & $1$ & * &$*$ & &  $2999.39$ &  &  &   &  &  & \\
178 & $a_1$ & $1$ &  & $1$ & * &$*$ & &  $3003.62$ &  &  &   &  &  & \\
179/180 & $a_1/b_1$ &  & $2$ & $1$ & $2$ & $2^2 \cdot3^14^1$ & $2^2 \cdot$VI: $1/3$ ($a_1'+a_2'+\underline{e'}$) & $3005.36$ &  &  &   &  &  & \\
\bottomrule
\end{longtable}
\endgroup
\end{landscape}
\if\USETWOCOLUMNS1
\twocolumn
\fi

Compared to the previously computed levels,
our energies differ greatly from that of \lit{NO32018viel} with a discrepancy of up to $\unit[126]{\icm}$ for state 26 ($4^4$). This discrepancy can be understood in terms of a less accurate vibronic potential energy surface used in \lit{NO32018viel}, compared to the one used here.\cite{Accurate2021viel}
A slightly improved agreement can be found between our fully \textit{ab initio} levels and the semi-empirical ones from \lit{Communication2014homayoon}, where the maximal discrepancy is  $\unit[100]{\icm}$ for state 16 ($3^1 4^1$). %
The computed levels from \lit{Simulation2022stanton}, which uses the Hamiltonian from \lit{N32012simmons}, have served as reference in a multitude of experimental studies, see, e.g., \lits{Fourier2013fujimori,Spectroscopic2013takematsu,Photoelectron2015gozema,Jet2015codd,HighResolution2020babin}.
Our levels are in good agreement with those from \lit{Simulation2022stanton}.
Compared to our energies, the largest deviation is $\unit[43]{\icm}$ for states 52/53 ($4^5$).
The good agreement is remarkable, given the many challenges that are associated with computing the electronic structure, non-adiabatic couplings, and fitting the non-adiabatic potential energies for this molecule.
Note that the potential fit used in this work has not been designed to give accurate vibronic states of a particular adiabatic state but to accurately reproduce the photodetachment spectrum of \ce{NO3-}, which requires a balanced description of all five electronic states involved across a large part of configuration space.\cite{Neural2018williams,Accurate2021viel,Simulation2022williams}

While comparing levels computed using different theories is useful, a comparison to experimental values is even more illuminating and, in this case, may also reveal discrepancies in the experimental assignment. 
Overall, the agreement of our computed levels with experiment is very good. 
Computed energies for states with excitations in $Q_1$ or $Q_2$ are systematically larger than the experimental references, whereas those for excitations in $Q_3$ or $Q_4$ are smaller  but closer to the experimental references. Note that two levels have been reported in a  Laser Induced Fluorescence (LIF) spectroscopy experiment by Kim et al.\cite{NO31992kim} that are below $\unit[3000]{\icm}$ and not included here, as these levels belong to excitation groups that are not fully included in our list of assigned states in \autoref{tab:assignment}. An assignment of an even larger number of computed states would be required for a proper analysis of these two peaks.
Assuming that the experimental assignment is correct (see below), the largest deviation to experiment is for the $3^1 4^2$ states 31/32 with a deviation of $\sim\unit[180]{\icm}$. 
The RMSD for all levels is $\unit[58]{\icm}$, which includes uncertain experimental assignments (see below).

There are five levels with differences of  more than  $\unit[100]{\icm}$ compared to reported levels from experimental gas phase studies. All five levels have excitations in $Q_4$.
The IR inactive state 15 ($4^3$) with $a_1'$ symmetry has only recently been detected through LIF spectroscopy,\cite{Laser2022fukushima} where an excitation energy corresponding to $\unit[1055]{\icm}$ has been reported.
This value is in disagreement with not only our computed energy of $\unit[1225]{\icm}$, but also with the three other, independent theoretical predictions.
State 16 ($3^1 4^1$) is difficult to detect experimentally due to a weak intensity.\cite{FTIR2013kawaguchi}
As it could not be directly detected, the reported\cite{Infrared2017kawaguchi} experimental value of $\unit[1491(3)]{\icm}$ was determined through perturbation analysis.\cite{FTIR2013kawaguchi}
Our computed value is $\unit[1330]{\icm}$, which, as state 15, is close to previous theoretical predictions.\cite{NO32018viel,Diabatic2019williams,Simulation2022stanton}
All other levels with $3^1 4^1$ excitations are in very good agreement with experiment.

{A similar pattern as for $3^1 4^1$ exists for $3^1 4^2$.
All excitations of type $3^1 4^2$ but one are in very good agreement with experiment.
With a deviation of $\unit[186]{\icm}$,
states 31/32 ($3^1 4^2$) have the largest disagreement to experiment but, like $3^1 4^1$, the experimental value of $\unit[1950]{\icm}$ is not directly observed but stems from a perturbation analysis.}
The same issue about deviations for these two excitations with very weak IR intensity has been discussed in \lit{Simulation2022stanton}. 
Our values are in very good agreement with \lit{Simulation2022stanton}.
Additional computations on isotopologues, where some transitions have larger IR intensities,\cite{Infrared2017kawaguchi} would help to illuminate more aspects of this, but this is beyond the scope of this work.
The other two states with energies that have large disagreement with a LIF experiment\cite{Laser2022fukushima} are state 74 ($1^1 4^3$) and state 117 ($1^1 4^4$). 
To summarize, three of the five levels with large disagreement to experiment are based on a recent LIF experiment,\cite{Laser2022fukushima}
and the other two levels with large disagreement %
{have a very weak IR intensity and the reported experimental values are based on perturbation analysis rather than direct observation.}
Other levels  of the same type of excitation patterns with stronger intensity have excellent agreement with experiment.
More experiments and computations using different potentials and targeting also high-lying eigenstates
may help to clarify the disagreement.

As previous computational studies,\cite{Vibronic2007stanton,N32012simmons,Communication2014homayoon,NO32018viel}
we also confirm the re-assignment of states 7/8 ($3^1$) from  $\unit[1492]{\icm}$ 
to the experimental transition at $\unit[1054]{\icm}$,
and states 19/20 ($3^1 4^1$) to the transition at $\unit[1492]{\icm}$.\cite{N32012simmons}
Likewise, we confirm the recently reported assignment of state 26 ($4^4$ with $e'$ symmetry)
to the transition at $\unit[1567]{\icm}$ measured using photodetachment spectroscopy.\cite{HighResolution2020babin}
A transition at $\unit[1608]{\icm}$ recently reported for this state through LIF experiments\cite{Laser2022fukushima}
would much better match the energy of state 26 ($4^4$ with $a_1'$ symmetry, calculated as $\unit[1587]{\icm}$), which is in agreement with the results and discussion in \lit{Simulation2022stanton}.

In the experimental IR spectrum, there are hot bands arising from the low-energy state 1 ($4^1$) that recently have been analyzed in detail.\cite{Fourier2011kawaguchi,FTIR2013kawaguchi,Infrared2017kawaguchi}
In particular, 
there are  possible transitions to two states of $a'$ symmetry with a $3^14^2$ excitation, but only one of them led to an observed hot band due to low intensity. Assuming validity of our dipole surface (see \autoref{subsec:IR_spec}), we can confirm this: 
The computed hot band IR intensity of state 41 is a factor of 6 lower than that of state 42.

The large spread of the $3^1 4^1$ and $3^1 4^2$ sublevels has been remarked in \lit{Simulation2022stanton},
and we can confirm this spread. 
However, as discussed above, more experiments would help to confirm or to disprove this spread of sublevels.
Also, we notice in our results a 
large level splitting of many other excitation groups such as $3^1 4^3$ and $3^2 4^1$. For these levels, an experimental verification is missing, save for states 83/84 and 108/109, whose energies are in excellent agreement with experiment.

In addition to the splitting of some combination bands, we observe a surprisingly large splitting of the levels with large excitations in $Q_4$: The first of the five $4^6$ levels is of $a_2'$ symmetry with an energy of $\unit[2042]{\icm}$. Remarkably, this level is sandwiched energetically by the $4^5$ states!
The next state belonging to the $4^6$ has an energy that is $\unit[439]{\icm}$ higher (state 91 with $a_1'$ symmetry).
This is a reoccurring pattern also for the $4^7$ and $4^8$ levels (compare states 106, 110, 167/168 and 169), and for combination bands (e.g.~states 143/144 and 145).
Thus, for high excitations in $Q_4$, the energy splitting imposed by symmetry or vibrational angular momentum is larger than the zero-order reference.

Next to these large splittings, we also observe possible resonances and a strong mixing of the $3^2 4^2$ and the $1^1 3^1 4^2$ levels (groups XVIII and $1^1\cdot$ VIII in \autoref{tab:assignment}). 
The harmonic energies of these two excitations are energetically close ($\unit[2776]{\icm}$ and $\unit[2811]{\icm}$), confirming the possibility for the occurrence of resonances. 
The grouped assignment in \autoref{tab:assignment} only displays the largest zero-order component and, due to the mixing, is parenthesized. 
Due to a similar type of mixing, some states belonging to group XIV are parenthesized. 
Note that for these cases the grouped assignment may be different from the  quanta in the individual modes given in \autoref{tab:assignment}. The grouped assignment does take additional symmetry and energy considerations into account.

The discussion about the large energy splittings and possible resonances indicate the difficulty of assigning states with  many excitations in $Q_3$ and/or $Q_4$.
In general, the assignment based on VSCF states is straightforward for excitations in $Q_1$, $Q_2$ (see below for a discussion), and for some lower excitations in $Q_4$.
In contrast,
for many excitations in $Q_3$ and $Q_4$, the assignment is not always clear and could only be done using a combination of observation of energies, symmetries, VSCF overlaps, and wavefunction cuts. 
The assignment was more difficult for high-energy states containing multiple vibrational quanta in each mode. Difficult assignments for such states is typical for strongly coupled, anharmonic systems.\cite{Vibronic2009stantona,Resonance2018larsson,Variational2020wang}
The zero-order picture drawn from the assignment thus must be taken with some caution.
For example, for many states the wavefunction cuts display more subtle excitation patterns. Often, lobes of small magnitude are visible in the cuts even for states formally marked as having no excitations in that mode.
This is particularly the case for high-energy states and cuts along $Q_3$ and $Q_4$, but also happens for $Q_1$ and partially $Q_2$ for states with energies larger than $\unit[2000]{\icm}$.
Note that the wavefunction cuts expose these subtle excitation patterns 
but not the reduced densities, which reflect the average excitation.

\begin{figure*}
\includegraphics[scale=.9]{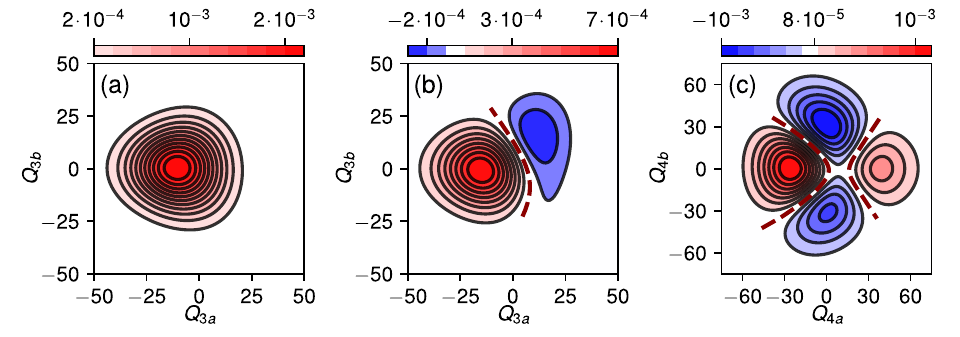}
  \caption{Wavefunction cuts for state 5 in \autoref{tab:assignment} ($4^2$).  
  Shown are two cuts along the $Q_3$ modes at different positions of $Q_{4}$ in panels (a) and (b), and cuts along the $Q_4$ modes in panel (c).
 }
  \label{fig:coupling_v3_v4}
\end{figure*}

To shine more light on the intriguing nature of the vibronic dynamics of \ce{NO3}, we now discuss the wavefunctions of some exemplary states.
A strong coupling of the antisymmetric stretching and bending coordinates $Q_3$ and $Q_4$ already is visible even for state 5, which is assigned as $4^2$ ($\ket{11}$ in the two $Q_4$ coordinates) but displays a weak excitation pattern along the $Q_3$ modes.
This is depicted in \autoref{fig:coupling_v3_v4}.
Note further the strong anharmonicity in all four modes.
In general, we observe that the higher the excitation in $Q_4$, the more coupling to the $Q_3$ modes there is.
This is not surprising as the two degenerate modes are not pure stretching and bending modes (compare with \autoref{fig:modes}),  as has been observed previously.\cite{Vibronic2007stanton,NO32018viel}
An alternative to the normal coordinates of the $\ce{NO3}$ radical is to use the normal coordinates of $\ce{NO3-}$ as reference,\cite{Ground1994mayer}
but Duchinsky rotations lead to a similarly large coupling in these coordinates.\cite{Vibronic2007stanton}
Exploring other coordinates for the assignment is beyond the scope of this work.

\begin{figure*}
\includegraphics[scale=.9]{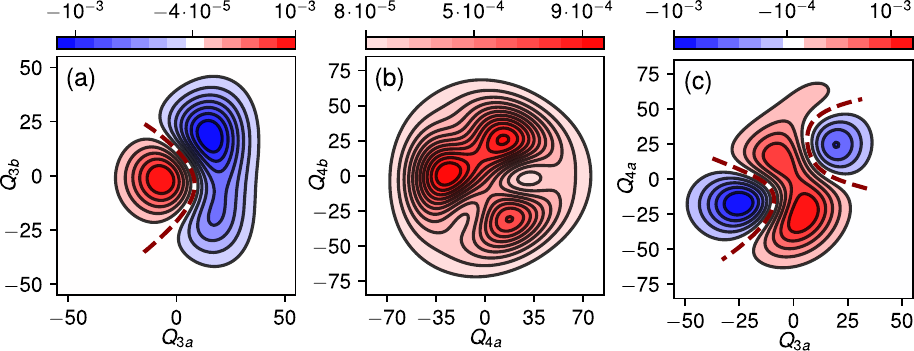}
  \caption{Wavefunction cuts for state 21 in \autoref{tab:assignment} ($3^1 4^1$). 
  Shown are
  cuts along $Q_3$ (a), $Q_4$ (b) and $Q_{3a}/Q_{4a}$ (c).
  Cuts along  $Q_{3b}/Q_{4b}$ are similar to those shown in panel (c).}
  \label{fig:mystery21}
\end{figure*}

Another large effect of correlation is shown in \autoref{fig:mystery21} for the IR active state 21.
While the state formally is assigned as $3^1 4^1$ excitation, some cuts in the $Q_4$ modes do not display any excitation patterns but rather an extended positive wavefunction lobe, see \autoref{fig:mystery21} (b).
The excitation in $Q_4$ only is revealed by inspecting plots along the $Q_3$ and $Q_4$ modes.
Singular value decomposition along the $Q_3$ modes and the $Q_4$ modes reveals three important states, 
the largest (!) component being
an approximate ground state of type $\ket{v_{Q_{3a}}v_{Q_{3b}}} \times 
\ket{v_{Q_{4a}}v_{Q_{4b}}} =  \ket{00} \times \ket{00}$, 
and the other two large components being of type
 $\ket{01} \times \ket{01}$, and $\ket{10}\times \ket{10}$, respectively.
 Indeed, a $\ket{v_{Q_{3a}}v_{Q_{4a}}}=\ket{11}$ type of excitation is visible in \autoref{fig:mystery21}(c).
This is a clear vibrational correlation effect and does not occur in VSCF computations. Only if all four modes $Q_{3a}, Q_{3b}, Q_{4a}$ and $Q_{4b}$ are correlated do we see the same excitation pattern.
Note that this type of pattern, in particular the extended lobe in $Q_4$ in \autoref{fig:mystery21}(b), is reoccurring. Specifically, it is visible in states 43/44, 73, 81, 99, 108/109, 127/128, 130, and 171/172.

\begin{figure}
\includegraphics[scale=.9]{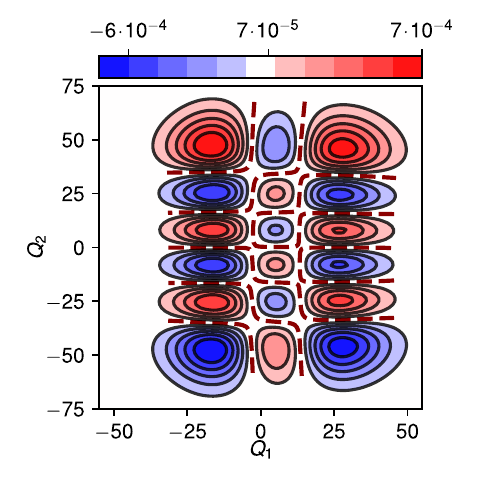}
  \caption{Wavefunction cut for state $1^2 2^5$ in \autoref{tab:fundamentals}.}
  \label{fig:coupling_v1_v2}
\end{figure}

In contrast to states with excitations in $Q_3$ or $Q_4$, states with excitations only in $Q_1$ or $Q_2$ are close to harmonic  and easy to assign. For example, \autoref{fig:coupling_v1_v2} displays a cut for the $1^22^5$ state. %
Despite the large excitations both in $Q_1$ and $Q_2$, this state displays a clear excitation pattern. Note, the strong displacement along $Q_1$, however.

States we found with excitations in $Q_1$ and $Q_2$ are tabulated in \autoref{tab:fundamentals},  and their energies are compared to harmonic estimates. 
States having only excitations in $Q_1$ mostly follow the harmonic approximation. There is a small negative anharmonicity 
due to vibronic interactions in $Q_3/Q_4$, but 
the difference to the harmonic estimate for $1^5$ only is $\unit[-24]{\icm}$.
There is a more pronounced positive anharmonicity for states with excitations in $Q_2$, with the maximum deviation to the harmonic estimate being $\unit[179]{\icm}$ for $2^7$. Due to coupling, combination bands have slightly larger deviations. 
The highest-energy state we found is $1^3 2^4$ with an energy deviating only $\unit[153]{\icm}$ from the harmonic estimate.

\begin{table}[h]
\small
  \caption{Fundamentals and combination bands of excitations in $Q_1$ and $Q_2$. Shown are the quantum numbers in $Q_1$,$Q_2$, computed term values, corresponding  harmonic term values (with zero-point energy taken from the computed energies), and difference of the harmonic and the computed term values.}
  \label{tab:fundamentals}
 \begin{tabular*}{.48\textwidth}{@{\extracolsep{\fill}}ccrrr}
  \toprule
  $v_1$ & $v_2$ & $\tilde E_{\text{calc.}}/\unit{\icm}$ &%
 $\tilde E_{\text{harm.}}/\unit{\icm}$ &%
  $\Delta/\unit{\icm}$ %
  \\
  \midrule
$0$ & 1 & $793$ & $793$ & $0$\\
$1$ & 0 & $1064$ & $1064$ & $0$\\
$0$ & 2 & $1543$ & $1586$ & $43$\\
$1$ & 1 & $1855$ & $1857$ & $2$\\
$2$ & 0 & $2138$ & $2129$ & $-9$\\
$0$ & 3 & $2318$ & $2379$ & $62$\\
$1$ & 2 & $2593$ & $2650$ & $58$\\
$2$ & 1 & $2926$ & $2922$ & $-4$\\
$0$ & 4 & $3080$ & $3172$ & $92$\\
$3$ & 0 & $3189$ & $3193$ & $4$\\
$1$ & 3 & $3367$ & $3443$ & $77$\\
$2$ & 2 & $3654$ & $3715$ & $61$\\
$0$ & 5 & $3845$ & $3965$ & $120$\\
$3$ & 1 & $3978$ & $3986$ & $8$\\
$1$ & 4 & $4125$ & $4236$ & $111$\\
$4$ & 0 & $4243$ & $4257$ & $14$\\
$2$ & 3 & $4422$ & $4508$ & $86$\\
$0$ & 6 & $4611$ & $4758$ & $147$\\
$3$ & 2 & $4695$ & $4779$ & $84$\\
$1$ & 5 & $4887$ & $5029$ & $143$\\
$4$ & 1 & $5033$ & $5050$ & $18$\\
$2$ & 4 & $5183$ & $5301$ & $118$\\
$5$ & 0 & $5345$ & $5322$ & $-24$\\
$0$ & 7 & $5372$ & $5551$ & $179$\\
$3$ & 3 & $5460$ & $5572$ & $112$\\
$1$ & 6 & $5653$ & $5823$ & $170$\\
$4$ & 2 & $5750$ & $5843$ & $94$\\
$2$ & 5 & $5934$ & $6094$ & $159$\\
$3$ & 4 & $6213$ & $6365$ & $153$\\
\bottomrule
   \end{tabular*}
\end{table}

\begin{figure*}
\includegraphics[scale=.9]{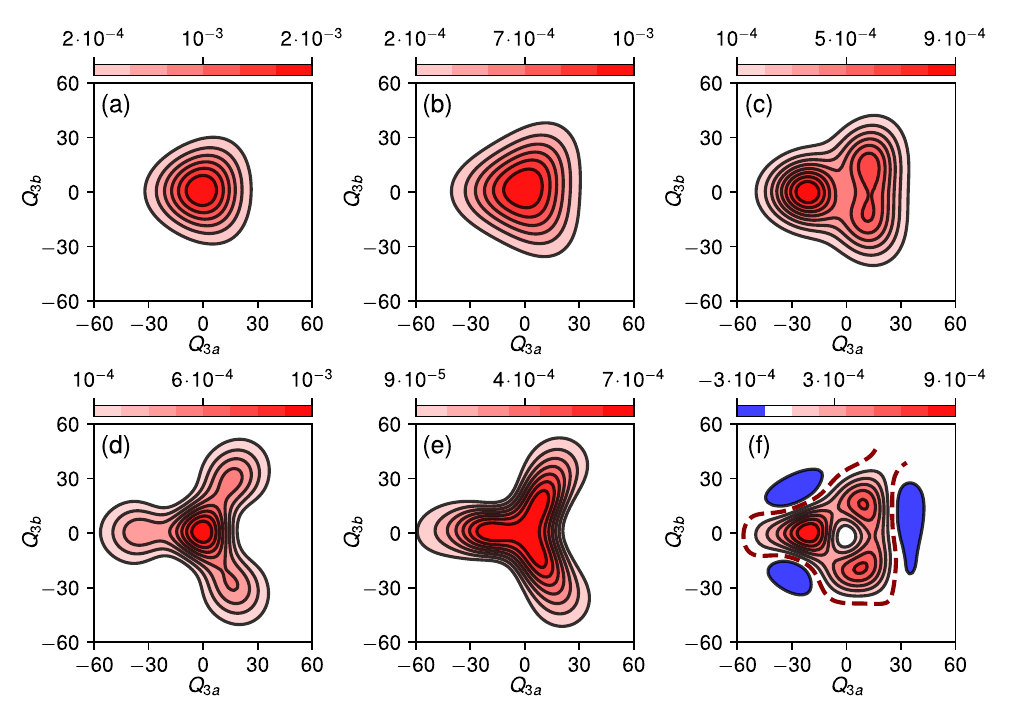}
  \caption{
  Wavefunction cuts along the $Q_3$ modes for the following states with excitations in $Q_1$: ground state (a), $1^1$ (b), $1^2$ (c),$1^3$ (d),$1^4$ (e), and $1^5$ (f). 
   Note that the potential along the cuts shown is triangular-shaped.
  Wavefunction cuts of $1^5$ display major distortions in $Q_4$ also (not shown).  }
  \label{fig:coupling_v1_v3}
\end{figure*}

While correlation between $Q_1$ and $Q_2$ is weak, there is a pronounced correlation between $Q_1$ and $Q_3$, as 
shown in \autoref{fig:coupling_v1_v3}.
The potential along $Q_3$ mostly is triangular in high-energy regions and distortions due to the multiwell structure of the wavefunction in $Q_3$  are visible for any node with excitations in $Q_1$.
A distinct multiwell character is first exposed for $1^2$ (state 59) shown in \autoref{fig:coupling_v1_v3}(c). 
The higher the excitation in $Q_1$, the more pronounced the multiwell structure in $Q_3$. 
This includes excitation patterns in $Q_3$ that are noticeable in the cuts. 
An example of this is visible in \autoref{fig:coupling_v1_v3}(f) for $1^5$. 
Due to the strong $Q_3-Q_4$ modal coupling, for these states excitation patterns are visible even in $Q_4$.
Despite these distortions in $Q_3$ and sometimes even $Q_4$, 
the excitation patterns in $Q_1$ remain close to harmonic with some more pronounced irregularities only visible for $1^5$.

\subsection{Infrared spectrum}
\label{subsec:IR_spec}
\begin{figure*}
\includegraphics[scale=.9]{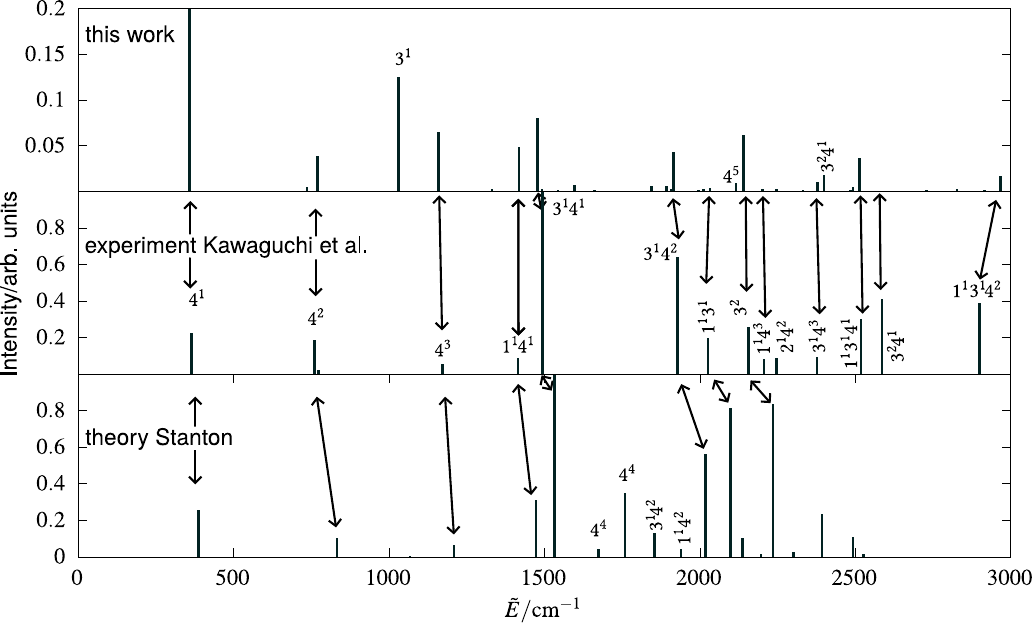}
  \caption{IR stick spectrum.
  Our computation based on non-adiabatic vibronic eigenstates 
  but a dipole moment surface for the lowest-energy state only   (upper panel), experiment (middle panel, as compiled in \lit{Infrared2018kawaguchi}), and IR spectrum  computed in \lit{Vibronic2009stantona} (lower panel). Note that the latter is based on vibrational eigenstates computed on an adiabatic Born-Oppenheimer PES without including the umbrella mode. %
  The assignment of the states is denoted by $Q^n$, which means that there are $n$ quanta in mode $Q$.
  }
  \label{fig:ir_spec}
\end{figure*}

\autoref{fig:ir_spec} displays
the IR stick spectrum up to $\tilde E = \unit[3000]{\icm}$ (the region in where all of our states have been assigned)
and compares it to experiment\cite{Infrared2018kawaguchi} and  a previously reported Born-Oppenheimer computation.\cite{Vibronic2009stantona}
As already shown in \autoref{subsec:assignment},
except for a few outliers the line positions from our computations are in very good agreement with experiment.
The IR intensities, however, differ vastly.\footnote{Note that there are two $3^2 4^1$ states with $e'$ symmetry, 
87/88 in 108/109 in \autoref{tab:assignment}. Only one of them has a strong IR intensity. Our IR spectrum predicts a large intensity for state 87/88 (vibrational angular momentum in $Q_3$ of $l_3=\pm 1$) whereas experiments predict a large intensity for state 108/109 with $l_3=0$.\cite{Infrared2018kawaguchi}}
For example, our spectrum contains a large intensity for the $3^1$ state whereas the experimental intensity of that state is very small.\cite{Infrared2021kawaguchi}
The large differences to the experimental IR intensities are in agreement with \lit{NO32018viel}, which introduced the DMS surface we used for computing the IR intensities. Note that this DMS only describes the adiabatic ground state and thus neglects the other electronic states.
In \lit{NO32018viel}, 
the differences of the intensities to experiment are explained by missing non-adiabatic effects. For example, the correct low intensity of the vibronic $3^1$ state might be from pseudo Jahn-Teller coupling of the $\tilde X ^2A_2'$ electronic state to the $\tilde B ^2E'$ state and intensity cancellation. \cite{FTIR2013kawaguchi} This cannot be described by the current DMS.
Stanton observed a similar effect for the intensities of the dispersed fluorescence spectrum when removing the pseudo Jahn-Teller coupling.\cite{Simulation2022stanton}
Note, however, that the IR spectrum from \lit{Vibronic2009stantona} that is displayed in the lower panel of \autoref{fig:ir_spec} is in better albeit not perfect agreement with experiment but also was computed by neglecting non-adiabatic effects. 
One explanation for the better agreement might be that the computation from \lit{Vibronic2009stantona} not only neglected non-adiabatic effects but also did not include the umbrella $Q_2$ mode. Since distortions along that mode leads to non-adiabatic coupling to the ground state and pseudo Jahn-Teller distortions, the relatively good intensities of  the IR spectrum from \lit{Vibronic2009stantona} might come from error cancellation (lack of both non-adiabatic effects and motion along the $Q_2$ mode).
As side remark, note that the energies from \lit{Simulation2022stanton} and shown in \autoref{tab:assignment}
are different to those from the same author reported in \lit{Vibronic2009stantona} and shown in \autoref{fig:ir_spec}.
The energies from \lit{Simulation2022stanton}  are based on a vibronic coupling Hamiltonian\cite{Multimode1984koppel} whereas the energies from \lit{Vibronic2009stantona} are based on an adiabatic force field.

To confirm that our poor IR intensities are not due to an improper description of the dipole moments of the adiabatic ground state, we generated another DMS.
Compared to the DMS fit of \lit{NO32018viel}, our DMS is based on interpolation and a different electronic structure method that is  similar to the one used in  \lit{Vibronic2009stantona}.
Nevertheless, also this new DMS results in a  similarly poor description of the IR intensities.
Along the same argument, the IR spectrum using the same DMS but eigenstates obtained using the Born-Oppenheimer approximation has much fewer intense peaks; see Section S1 and Figure S1 
in the ESI for details.
A generation of a DMS that takes all five electronic states into account is beyond the scope of this work. At any rate, our results show more evidence that non-adiabatic effects are important for correctly describing the IR spectrum.
The non-adiabatic effects onto the vibronic energies are 
discussed more in \autoref{sec:nonadiabatic_effects}.

\subsection{Non-adiabatic effects}
\label{sec:nonadiabatic_effects}

Some previous studies used a purely adiabatic Born-Oppenheimer PES 
for computing the vibrational states of \ce{NO3} for the  $\tilde X ^2A_2'$  ground state,\cite{Vibronic2009stantona,Communication2014homayoon,Diabatic2019williams}
whereas others, including this one, used a full non-Born-Oppenheimer diabatic PES model and thus compute vibronic states.\cite{Vibronic2007stanton,NO32018viel,Simulation2022stanton}
While there is a pseudo Jahn-Teller effect between the $\tilde X ^2A_2'$  diabatic state  and the $\tilde B ^2E'$ diabatic excited state,\cite{Ground1994mayer,Vibronic2007stanton}
this can be dealt with in a Born-Oppenheimer PES as long as the distortions are well-described by such a PES and as long as the non-adiabatic coupling in regions of large amplitude of the vibronic states is not large.
Then how important actually are non-Born-Oppenheimer effects?
\autoref{subsec:IR_spec} already hinted at the possibility that  they can significantly affect observables.

\begin{figure*}
\centering{
         \includegraphics[scale=1]{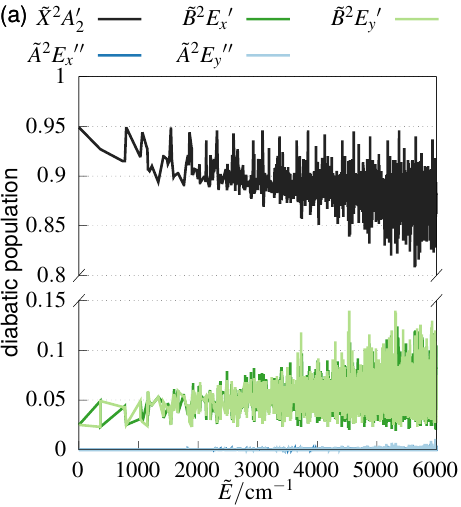}
         \includegraphics[scale=1]{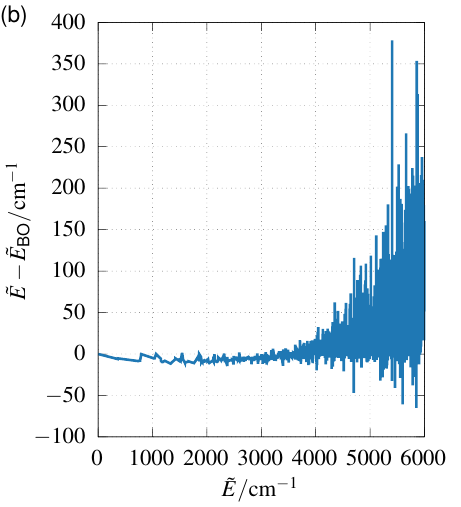}\\
         }
  \caption{Nonadiabatic effects: (a) Diabatic populations of the computed vibronic states as a function of their energy. Note the range cut in the ordinate axis. (b) Difference between vibronic energies and BO energies as a function of energy. BO energies were sorted based on the overlap of the BO states to the non-BO states.}
  \label{fig:bo_vs_nonbo}
\end{figure*}

To address the question of how important non-Born-Oppenheimer effects are, \autoref{fig:bo_vs_nonbo}  (a) displays the diabatic populations as function of energy.
Importantly, the diabatic ground state only has a population of $0.95$ even for the vibrational ground state. This comes from the diabatic model and the fact that the diabatic couplings (the off-diagonal part in \autoref{eq:vmatrix}) are non-zero,
save for configurations at the $D_{3h}$ geometry.
All vibronic wavefunctions spread out of $D_{3h}$ geometries. %
Starting with a population of $0.95$ for the ground state,
for excited states the population quickly decreases further to $\sim 0.9$ around $\unit[1000]{\icm}$. Populations of $\sim 0.85$ are reached around $\unit[4000]{\icm}$.
These populations are lower than those from Viel and Eisfeld,\cite{NO32018viel} who report populations of $0.96$ around $\unit[1000]{\icm}$, indicating stronger non-adiabatic effects described by our PES in comparison to the one used by Viel and Eisfeld known to have flaws.
With the ground state population slowly decreasing, 
the populations of the two $E'$ states are slowly increased whereas the populations of the $E''$ states remain negligible.
Since only distortions along the umbrella $Q_2$ motion lead to couplings to the $E''$ state, the negligible populations of these states is expected.

How much does the decrease in ground state population affect other observables? 
To estimate this, we conducted an independent computation on the adiabatic ground state surface only (i.e., a Born-Oppenheimer computation). The energetic errors resulting from the omission of non-Born-Oppenheimer effects are displayed in  \autoref{fig:bo_vs_nonbo} (b).  
The errors can be large and reach almost $\unit[400]{\icm}$ in the computed region.
This corresponds to a relative deviation of up to $7\%$, which is in close agreement with the populations shown in \autoref{fig:bo_vs_nonbo} (a).
Interestingly, the Born-Oppenheimer error first 
is negative at energies below $\sim\unit[4000]{\icm}$ but at higher excitation turns positive for most states.
For energies below $\unit[400]{\icm}$ the largest error is $\sim\unit[-14]{\icm}$.
As mentioned in \autoref{subsec:IR_spec},
the IR spectrum computed from the Born-Oppenheimer vibrational states displays much fewer non-negligible peaks compared to the IR spectrum computed using non-BO states shown in \autoref{fig:ir_spec}, highlighting further the importance of the inclusion of the excited electronic states. %

\section{Conclusions}
\label{sec:conclusions}

The \ce{NO3} radical features an intriguing vibronic structure where to date many aspects of it, including the infrared spectrum, remain elusive. 
Here, using a recently developed, accurate PES that captures all five important electronic states, curvilinear coordinates and the tree tensor network state (TTNS) approach, we analyzed the vibronic structure of \ce{NO3}.
Specifically, we  increased the number of previously computed vibronic states by a factor of $\sim 50$ to more than 2500. We are not aware of a larger number of vibronic states being computed for a six-dimensional, highly anharmonic and non-adiabatic molecule.
This shows the  effectiveness of our recently developed TTNS approach.
Furthermore, we  increased the number of previously assigned states by a factor of $\sim 5$ from $~\sim$35 to  180. All states up to $\unit[3000]{\icm}$ are now assigned.
Even though the used PES was not specifically designed for accurate vibronic eigenstate computations for the electronic ground state, overall we achieved very good agreement with experiment. 
In agreement with previous computations, our analysis hinted at the possibility of required reassignments of experimental levels.

Our analysis of the vibronic states revealed a  strong correlation not only between the antisymmetric stretch and bending modes, $Q_3$ and $Q_4$, but also between the symmetric and antisymmetric stretch modes, $Q_1$ and $Q_3$.
In addition, we found a large spread of sublevels with excitations in the antisymmetric bending mode $Q_4$, thus showing that the energy splitting imposed by symmetry can even be  larger than the zero-order reference. 
The higher the energy, the more difficult the assignment and hence the less useful the zero-order harmonic picture. 
We found evidence for resonances and significant mixing of zero-order states for energies above $\unit[2800]{\icm}$. 
Thus, an energy region up to $\unit[\sim 3000]{\icm}$ seems to be most suitable for an assignment, albeit we found straightforwardly assignable combination states  in less-coupled modes %
even at an energy of $\unit[6213]{\icm}$, which is just $\unit[850]{\icm}$ below the electronic $\tilde A ^2E''$ state.
Nonadiabatic effects are surprisingly strong and the  $\tilde B ^2 E'$ states contribute between 5 and 10\% to the vibronic eigenstates below  $\unit[3000]{\icm}$. The contribution approaches 20\% for energies around $\unit[6000]{\icm}$. These non-negligible contributions lead to nontrivial effects on the vibronic energies and vast changes in the IR spectrum.
We hope that this work will  stimulate more computational and experimental studies on this intriguing free radical.

\section*{Author Contributions}
H.R.L.~and A.V.~conceived the study; 
H.R.L.~designed the study, developed the MPO/TTNS/DP-DVR methodology, performed and analyzed the TTNS and DP-DVR simulations;
A.V.~provided the Hamiltonian;
H.R.L.~and A.V.~analyzed the results.
H.R.L.~drafted the manuscript;
H.R.L.~and A.V.~edited the manuscript.

\section*{Conflicts of interest}
There are no conflicts to declare.

\section*{Acknowledgements}
H.R.L.~thanks A.~M.~Nierenberg for helpful discussions on coordinate transformations.
We thank W.~Eisfeld for providing us with the dipole moment surface.
Work by H.R.L.~was supported by the US
National Science Foundation (NSF) via grant no.~CHE-2312005.
Additional support came from University of California, Merced, start-up funding, 
and through  computational time on the Pinnacles and Merced clusters at University of California, Merced (supported by NSF OAC-2019144 and ACI-1429783).
A.V.~acknowledges partial financial support from the CNRS and the University of Rennes via the IRN MCTDH grant project, as well as from the French ANR agency under grant ANR-23-CE04-0014 (project RADICALS).

\balance

\renewcommand\refname{References}

\if\USEARXIV1
\bibliographystyle{apsrev4-1} 
\else
\bibliographystyle{rsc} 
\fi

\end{document}